\documentclass[11pt]{article}
\usepackage[utf8]{inputenc}
\usepackage{epsfig}

\usepackage[letterpaper]{geometry}
\usepackage{graphicx}
\usepackage[usenames,dvipsnames]{color}
\usepackage{todonotes}

\newcommand{\eps}{\varepsilon}

\newcommand{\R}{{\mathbb R}}
\newcommand{\N}{{\mathbb N}}

\newcommand{\dist}{\mathrm{dist}}
\newcommand{\emb}{\mathcal E}

\usepackage[numbers]{natbib}
\setcitestyle{aysep={}}

\usepackage{graphicx}
\usepackage[colorlinks=true, allcolors=blue]{hyperref}

\usepackage{paralist}
\usepackage{amsmath}
\usepackage{amssymb}
\usepackage{amsthm}
\usepackage{algorithmic}
\usepackage{algorithm}
\usepackage{tikz}

\DeclareMathOperator*{\argmin}{argmin}

\author{
Bogumi\l{} Kami\'nski\thanks{Decision Analysis and Support Unit, SGH Warsaw School of Economics, Warsaw, Poland; e-mail: \texttt{bogumil.kaminski@sgh.waw.pl}}
\and
Pawe\l{}~Pra\l{}at\thanks{Department of Mathematics, Ryerson University, Toronto, ON, Canada; e-mail: \texttt{pralat@ryerson.ca}}
\and
Fran\c{c}ois Th\'eberge\thanks{Tutte Institute for Mathematics and Computing, Ottawa, ON, Canada; email: \texttt{theberge@ieee.org}}
}

\title{An Unsupervised Framework for Comparing Graph Embeddings}

\begin{document}

\maketitle

\begin{abstract}
Graph embedding is a transformation of vertices of a graph into set of vectors. Good embeddings should capture the graph topology, vertex-to-vertex relationship, and other relevant information about graphs, subgraphs, and vertices. If these objectives are achieved, they are meaningful, understandable, and compressed representations of networks. They also provide more options and tools for data scientists as machine learning on graphs is still quite limited. Finally, vector operations are simpler and faster than comparable operations on graphs.

The main challenge is that one needs to make sure that embeddings well describe the properties of the graphs. In particular, the decision has to be made on the embedding dimensionality which highly impacts the quality of an embedding. As a result, selecting the best embedding is a challenging task and very often requires domain experts.

In this paper, we propose a ``divergence score'' that can be assign to various embeddings to distinguish good ones from bad ones. This general framework provides a tool for an unsupervised graph embedding comparison. In order to achieve it, we needed to generalize the well-known Chung-Lu model to incorporate geometry which is interesting on its own rights. In order to test our framework, we did a number of experiments with synthetic networks as well as real-world networks, and various embedding algorithms.
\end{abstract}

\section{Introduction}

The study of networks has emerged in diverse disciplines as a means of analyzing complex relational data. Indeed, capturing aspects of a complex system as a graph can bring physical insights and predictive power~\cite{Newman_book}. Network Geometry is a rapidly developing approach in Network Science~\cite{Bianconi2015} which further abstracts the system by modelling the vertices of the network as points in a geometric space. There are many successful examples of this approach that include latent space models~\cite{Hoff2002}, and connections between geometry and network clustering and community structure~\cite{Krioukov2016,Zuev2015}. Very often, these geometric embeddings naturally correspond to physical space, such as when modelling wireless networks or when networks are embedded in some geographic space~\cite{Gastner2006, Expert2011}. See~\cite{Janssen2010} for more details about applying spatial graphs to model complex networks.

Another important application of geometric graphs is in graph embedding. The idea here is that, for a given network, one tries to embed it in a geometric space by assigning coordinates to each vertex such that nearby vertices are more likely to share an edge than those far from each other. In a good embedding most of the network's edges can be predicted from the coordinates of the vertices. For example, in~\cite{Higham2008} protein interaction networks are embedded in low-dimension Euclidean space. Unfortunately, in the absence of a general-purpose representation for graphs, very often graph embedding requires domain experts to craft features or to use specialized feature selection algorithms. Having said that, there are some graph embedding algorithms that work without any prior or additional information other than graph structure. However, these are randomized algorithms that are usually not so stable; that is, the outcome is often drastically different despite the fact that all the algorithm parameters remain the same.

\medskip

Consider a graph $G=(V,E)$ on $n$ vertices, and several embeddings of its vertices to some multidimensional spaces (possibly in different dimensions). The main question we try to answer in this paper is: how do we evaluate these embeddings? Which one is the best and should be used? In order to answer these questions, we propose a general framework that assigns the divergence score to each embedding which, in an unsupervised learning fashion, distinguishes good from bad embeddings. In order to benchmark embeddings, we generalize well-known Chung-Lu random graph model to incorporate geometry. The model is interesting on its own and should be useful for many other problems and tools. In order to test our algorithm, we experiment with synthetic networks as well as real-world networks, and various embedding algorithms.

\medskip

The paper is structured as follows. In Section~\ref{sec:framework}, we describe our algorithm for comparing graph embeddings, and we illustrate our approach on one simple graph. The Chung-Lu model is generalized in Section~\ref{sec:chung-lu}. Some theoretical results that justify the model can be found in Appendix~\ref{apdx:model}. In Section~\ref{sec:experiments}, we describe the experiments and present results. In particular, the datasets and embedding algorithms used are outlined respectively in subsections~\ref{sec:exp_data} and \ref{sec:exp_emb}. We conclude with a discussion on some future directions in Section~\ref{sec:conclusion}.

\section{General Framework}\label{sec:framework}

Suppose that we are given a graph $G=(V,E)$ on $n$ vertices with the degree distribution $\textbf{w}=(w_1, \ldots, w_n)$ and an embedding of its vertices to $k$-dimensional space, $\emb : V \to \R^k$. Our goal is to assign a ``divergence score'' to this embedding. The lower the score, the better the embedding is. This will allow us to compare several embeddings, possibly in different dimensions.

\subsection{Intuition Behind the Algorithm}

What do we expect from a good embedding? As already mentioned, in a good embedding, one should be able to predict most of the network's edges from the coordinates of the vertices. Formally, it is natural to expect that if two vertices, say $u$ and $v$, are far away from each other (that is, $\dist(\emb(u),\emb(v))$ is relatively large), then the chance they are adjacent is smaller compared to another pair of vertices that are close to each other. But, of course, in any real-world network there are some sporadic long edges and some vertices that are close to each other are not adjacent. In other words, we do not want to pay attention to local properties such as existence of particular edges (microscopic point of view) but rather evaluate some global properties such as density of some relatively large subsets of vertices (macroscopic point of view). So, how can we evaluate if the global structure is consistent with our expectations and intuition without considering individual pairs?

The idea is as follows. We identify dense parts of the graph by running some good graph clustering algorithm. As we will illustrate in Section \ref{sec:experiments}, the choice of graph clustering algorithm is flexible so long as the vertex set is partitioned into clusters such that there are substantially more edges captured within clusters than between them. The clusters that are found will provide the desired macroscopic point of view of the graph.
Note that for this task we only use information about graph $G$; in particular, we do not use the embedding $\emb$ at all.
We then consider the graph $G$ from a different point of view.
Using the Geometric Chung-Lu (GCL) model that we introduce in this paper especially for this purpose, based on the degree distribution $\textbf{w}$ and the embedding $\emb$, we compute the expected number of edges within each cluster found earlier, as well as between them.
The embedding is scored by computing a divergence score between those expected number of edges, and the actual number of edges present in $G$.
Our approach falls into a general and commonly used method of \emph{statistical inference}, in our case applied to the Geometric Chung-Lu model. With these methods, one fits a generative model of a network to observed network data, and the parameters of the fit tell us about the structure of the network in much the same way that fitting a straight line through a set of data points tells us about their slope.

\subsection{Algorithm}

Given a graph $G=(V,E)$, its degree distribution $\textbf{w}$ on $V$, and an embedding $\emb : V \to \R^k$ of its vertices in $k$-dimensional space, we perform the five steps detailed below
to obtain $\Delta_\emb(G)$, a \emph{divergence score} for the embedding.
We can apply this algorithm to compare several embeddings $\emb_1,\dots,\emb_m$, and select the best one via $\argmin_{i\in[m]} \Delta_{\emb_i}(G)$ (here and later in the paper, we use $[n]$ to denote the set of natural numbers less than or equal to $n$; that is, $[n] := \{1, \ldots, n\}$). Note that our algorithm is a general framework and some parts have flexibility. We clearly identify these below.

\medskip \noindent \textbf{Step 1:} Run some stable \emph{graph} clustering algorithm on $G$ to obtain a partition $\textbf{C}$ of the vertex set $V$ into $\ell$ communities $C_1, \ldots, C_\ell$. \\
\noindent {\it Note:}
In our implementation, we used the ensemble clustering algorithm for graphs (ECG) which is based on the Louvain algorithm and the concept of consensus clustering~\cite{Poulin2019}, and is shown to have good stability.
We experiment with other algorithms in Section~\ref{sec:experiments}. \\

\medskip \noindent \textbf{Step 2:} For each $i \in [\ell]$, let $c_{i}$ be the proportion of edges of $G$ with both endpoints in $C_i$. Similarly, for each $1 \le i < j \le \ell$, let $c_{i,j}$ be the proportion of edges of $G$ with one endpoint in $C_i$ and the other one in $C_j$. Let
\begin{equation}
\bar{\textbf{c}} = (c_{1,2},\ldots, c_{1,\ell}, c_{2,3}, \ldots, c_{2,\ell}, \ldots, c_{\ell-1,\ell} )
\quad \text{and }
\hat{\textbf{c}} = (c_1, \ldots, c_\ell)
\label{eq:c}
\end{equation}
be two vectors with a total of $\binom{\ell}{2} + \ell = \binom{\ell+1}{2}$ entries that sum to one. These {\it graph vectors} characterize partition $\textbf{C}$ from the perspective of $G$. \\
\noindent{\it Note:}
The embedding $\emb$ does \emph{not} affect the vectors $\bar{\textbf{c}}$ and $\hat{\textbf{c}}$.
It is calculated purely based on $G$ and the partition $\textbf{C}$.

\medskip \noindent \textbf{Step 3:} For a given parameter $\alpha \in \R_+$ and the same vertex partition $\textbf{C}$, we consider $\mathcal{G}(\textbf{w}, \emb, \alpha)$, the GCL Model presented in Section~\ref{sec:chung-lu}. For each $1 \le i < j \le \ell$, we compute $b_{i,j}$, the expected proportion of edges of $\mathcal{G}(\textbf{w}, \emb, \alpha)$ with one endpoint in $C_i$ and the other one in $C_j$. Similarly, for each $i \in [\ell]$, let $b_i$ be the expected proportion of edges within $C_i$. That gives us another two vectors
\begin{equation}
\bar{\textbf{b}}_\emb(\alpha) = (b_{1,2},\ldots, b_{1,\ell}, b_{2,3}, \ldots, b_{2,\ell}, \ldots, b_{\ell-1,\ell} )  \qquad \text{ and } \qquad
\hat{\textbf{b}}_\emb(\alpha) = (b_{1},\ldots,  b_{\ell} )
\label{eq:b}
\end{equation}
with a total of $\binom{\ell+1}{2}$ entries that sum to one. These vectors characterizes partition $\textbf{C}$ from the perspective of the embedding $\emb$. \\
\noindent {\it Note:}
The structure of graph $G$ does \emph{not} affect the vectors $\bar{\textbf{b}}_\emb(\alpha)$ and $\hat{\textbf{b}}_\emb(\alpha)$; only its degree distribution $\textbf{w}$ and embedding $\emb$ are used.\\
\noindent {\it Note:}
We used the Geometric Chung-Lu Model but the framework is flexible. If, for any reason (perhaps there are some restrictions for the maximum edge length; such restrictions are often present in, for example, wireless networks) it makes more sense to use some other model of random geometric graphs, it can be easily implemented here. If the model is too complicated and computing the expected number of edges between two parts is challenging, then it can be approximated easily via simulations.

\medskip \noindent \textbf{Step 4:} Compute the distances between the two pairs of vectors, between $\bar{\textbf{c}}$ and $\bar{\textbf{b}}_\emb(\alpha)$, and between $\hat{\textbf{c}}$ and $\hat{\textbf{b}}_\emb(\alpha)$, in order to measure how well the graph $G$ fits the model $\mathcal{G}(\textbf{w}, \emb, \alpha)$. Let $\Delta_\alpha$ be a weighted average of the two distances. \\
\noindent {\it Note:} We used the well-known and widely used Jensen–Shannon divergence (JSD) to measure the dissimilarity between two probability distributions. The JSD was originally proposed in~\cite{Lin1991} and can be viewed as a smoothed version of the Kullback-Leibler divergence. In our implementation, we used simple average, that is,
\begin{equation}
\Delta_\alpha = \frac 12 \cdot \left(
JSD(\bar{\bf c},\bar{\bf b}(\alpha)) +  JSD(\hat{\bf c},\hat{\bf b}(\alpha))
\right).
\label{eq:delta}
\end{equation}
Other weighted averages can be used if more weight needs to be put on internal or external edges.

\medskip \noindent \textbf{Step 5:} Select $\hat{\alpha} = \argmin_{\alpha} \Delta_\alpha$, and define the \emph{divergence score} for embedding $\emb$ on $G$ as: $\Delta_\emb(G) = \Delta_{\hat{\alpha}}$. \\
\noindent \textbf{\it Note:}
The parameter $\alpha$ is used to define a distance in the embedding space, as we detail in Section~\ref{sec:chung-lu}.
In our implementation we simply checked values of $\alpha$ on a grid between 0 and 10. There are clearly better ways to search the space of possible values of $\alpha$ but, since the algorithm worked very fast on our graphs, we did not optimize this part.


\medskip

In order to compare several embeddings for the same graph $G$, we repeat steps 3-5 above and compare the divergence scores (the lower score, the better). Let us stress again that steps 1-2 are done only once, so we use the same partition into $\ell$ communities for each embedding.

\subsection{Illustration}

We illustrate our framework on the well-known Zachary's Karate Club graph (see Subsection~\ref{sec:exp_data} for the description of this and other datasets). Illustrations with other datasets are shown in Appendix~\ref{apdx:plots}.

\begin{figure}[ht]
\begin{center}
\includegraphics[width=7.5cm]{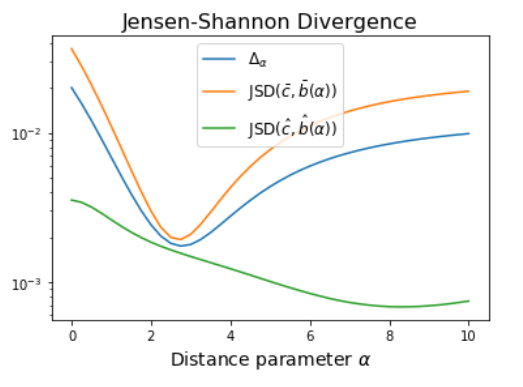}
\includegraphics[width=7.9cm]{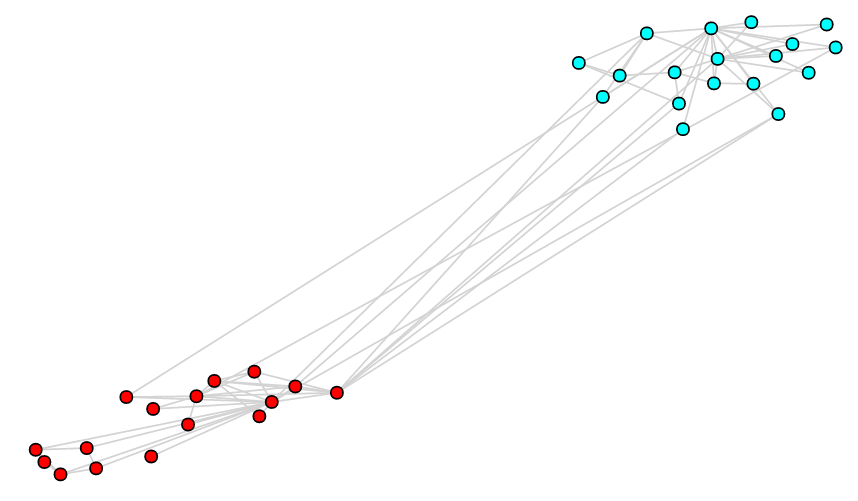}
\end{center}
\caption{The Karate Club Graph. We illustrate the divergence score as a function of $\alpha$ (left) for the best embedding found by our framework (right). The colors represent the two ground-truth communities.
\label{fig:karate-alpha1}}
\end{figure}

The parameter $\alpha \ge 0$ in the GCL model controls the distance used in the embedding space. With  $\alpha=0$, the embedding is \emph{not} taken into account and the classic Chung-Lu model is obtained, so only the degree distribution is accounted for.
As $\alpha$ gets larger, long edges in the embedding space are penalized more severely.
In the left plot of Figure~\ref{fig:karate-alpha1}, we show the impact of varying $\alpha$
on the two components of equation~(\ref{eq:delta}) which respectively consider pairs of vertices that are {\it internal} (to some cluster) or {\it external} (between clusters).
Recall that the divergence score for a given embedding is obtained by choosing
$\hat{\alpha} = \argmin_{\alpha} \Delta_\alpha$.
In the right plot of Figure~\ref{fig:karate-alpha1}, we show a 2-dimensional projection of
the best embedding as obtained by our framework. The vertices are coloured according to the
two known communities.

\begin{figure}
\begin{center}
\includegraphics[width=5.2cm]{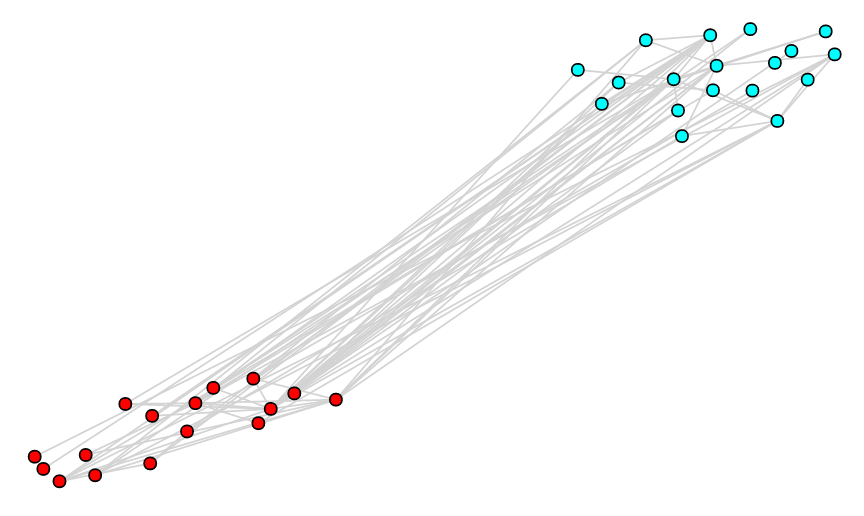}
\includegraphics[width=5.2cm]{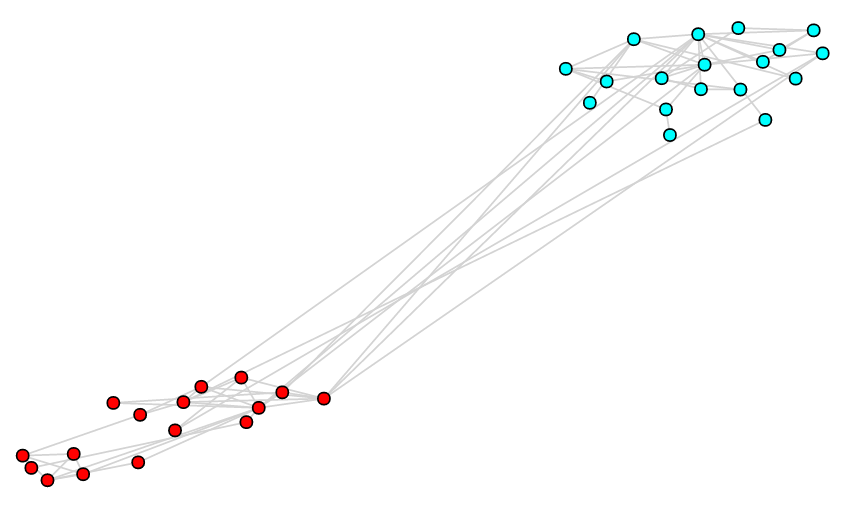}
\includegraphics[width=5.2cm]{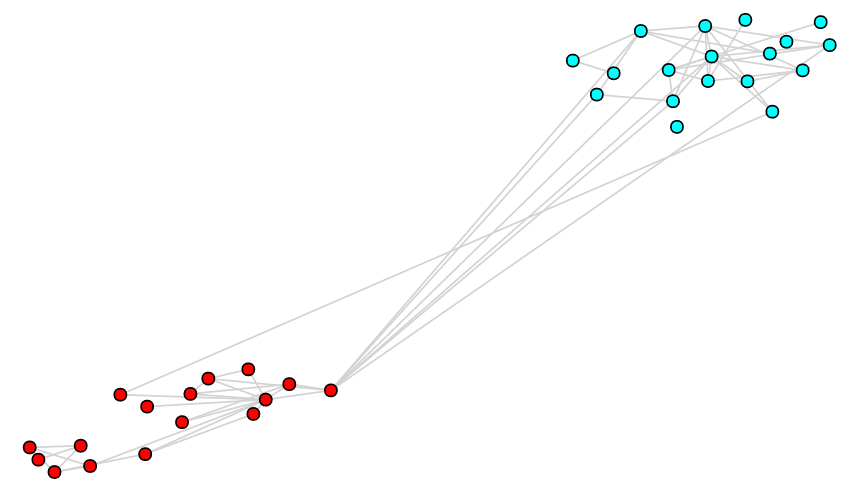}
\end{center}
\caption{Zachary's Karate Club Graph. We generate random edges following the Geometric Chung-Lu Model with the same expected degree distribution and with the highest scoring embedding. We look at three cases: $\alpha=0$ which ignores the embedding (left), $\alpha=7$ which penalizes long edges too severely (right), and the best $\hat{\alpha}=4$ (center).
\label{fig:karate-alpha2}}
\end{figure}

We can use the GCL model to generate edges, as with the standard Chung-Lu model.
In Figure~\ref{fig:karate-alpha2}, we generate 3 such graphs using the best embedding
shown in Figure~\ref{fig:karate-alpha1}. The left plot uses $\alpha=0$, which ignores the embedding and clearly generates too many long edges between the clusters.
The center plot uses the optimal value ($\hat{\alpha}=2.75$ in this case), generating a graph that resembles the true one. The rightmost plot uses the larger value $\alpha = 7$, which penalizes
long edges more severely, yielding a graph with less edges between the two communities.

\section{Geometric Chung-Lu Model}\label{sec:chung-lu}

It is known that classical Erd\H{o}s-R\'enyi (binomial) random graphs $G(n,p)$ can be generalized to $G(\textbf{w})$, the random graph with a given expected degree distribution $\textbf{w}=(w_1, \ldots, w_n)$. Because of our application, we will define it as a function of a given graph $G$ on $n$ vertices but, in fact, it is only a function of its degree sequence.

Since our goal is to compare different embeddings of the same graph, we will generalize the Chung-Lu model further, including geometry coming from graph embedding. In such models vertices are embedded in some metric space and link formation is influenced by the metric distance between vertices. Such models are called \emph{spatial models} or \emph{geometric graphs}. The main principle of spatial models is that vertices that are metrically close are more likely to link to each other. This is a formal expression of the intuitive notion we have about virtual networks: Web links are likely to point to similar pages, people that share similar interests are more likely to become friends on Facebook, and scientific papers mostly refer to papers on similar topics. See~\cite{Janssen2010} for more details about applying spatial graphs to model complex networks.

\subsection{Original Model}

Let $G=(V,E)$ be a graph, where $V = \{v_1,\ldots ,v_n\}$ are the vertices, the edges $E$ are multisets of $V$ of cardinality 2 (loops are allowed), and $\deg_G(v)$ is the degree of $v$ in $G$ (with a loop at $v$ contributing 2 to the degree of $v$).
We define $\mathcal{G}(G)$ to be the probability distribution on graphs on the vertex set $V$ following the well-known Chung-Lu model~\cite{CL, Seshadhri2012, Kolda2014, Winlaw2015}. In this model, each set $e=\{v_i,v_j\}$, $v_i,v_j \in V$, is independently sampled as an edge with  probability given by:
\[
P(v_i,v_j) =
\begin{cases}
\frac{\deg_G(v_i) \deg_G(v_j)}{2|E|}, & i \ne j \\
\frac{\deg_G^2(v_i)}{4|E|}, & i = j.
\end{cases}
\]
(Let us mention about one technical assumption. Note that it might happen that $P(v_i,v_j)$ is greater than one and so it should really be regarded as the expected number of edges between $i$ and $j$; for example, as suggested in~\cite{Newman_book}, one can introduce a Poisson-distributed number of edges with mean $P(v_i,v_j)$ between each pair of vertices $i$, $j$. However, since typically the maximum degree $D$ satisfies $D^2 \le 2 |E|$ it rarely creates a problem and so we may assume that $P(v_i,v_j) \le 1$ for all pairs.)

As already mentioned, this model is a function of the degree sequence of $G$. One desired property of this random model is that it yields a distribution that preserves the expected degree for each vertex, namely: for any $i \in [n]$,
\begin{eqnarray*}
\mathbb{E}_{G' \sim \mathcal{G}(G)}[\deg_{G'}(v_i)] &=& \sum_{j \in [n] \setminus \{i\}} \frac{\deg_G(v_i)\deg_G(v_j)}{2|E|} + 2 \cdot \frac{\deg_G^2(v_i)}{4|E|} \\
&=& \frac{\deg_G(v_i)}{2|E|} \sum_{j \in [n]} \deg_G(v_j) ~~=~~ \deg_G(v_i).
\end{eqnarray*}

\subsection{Geometric Model}

This time we are not only given the expected degree distribution
$$
\textbf{w}=(w_1, \ldots, w_n) = (\deg_G(v_1), \ldots, \deg_G(v_n))
$$
but also an embedding of its vertices in some $k$-dimensional space, function $\emb : V \to \R^k$. In particular, for each pair of vertices, $v_i$, $v_j$, we know the distance between them:
$$
d_{i,j} = \dist( \emb(v_i), \emb(v_j)).
$$
It is desired for the probability that $v_i$ and $v_j$ are adjacent to be a function of $d_{i,j}$, that is, to be proportional to $g(d_{i,j})$ for some function $g$. Function $g$ should be a decreasing function as long edges should occur less frequently than short ones. There are many natural choices such as $g(d) = d^{-\beta}$ for some $\beta \in [0, \infty)$ or $g(d) = \exp(-\gamma d)$ for some $\gamma \in [0, \infty)$. We use the following, normalized function $g:[0,\infty) \to [0,1]$: for a fixed $\alpha \in [0,\infty)$, let
$$
g(d) := \left( 1 - \frac{d - d_{\min}}{d_{\max} - d_{\min}} \right)^{\alpha},
$$
where
\begin{eqnarray*}
d_{\min} &=& \min \{ \dist(\emb(v), \emb(w)): v,w \in V \} \\
d_{\max} &=& \max \{ \dist(\emb(v), \emb(w)): v,w \in V \}
\end{eqnarray*}
are the minimum, respectively maximum, distance between vertices in embedding $\emb$. One convenient and desired property of this function is that it is invariant with respect to an affine transformation of the distance measure. Clearly, $g(d_{\min})=1$ and $g(d_{\max})=0$; in the computations, we can use clipping to force $g(d_{\min})<1$
and/or $g(d_{\max})>0$ if required.
Let us also note that if $\alpha = 0$ (that is, $g(d)=1$ for any $d \in [0,\infty)$ with $g(d_{\max})=0^0=1$), then we recover the original Chung-Lu model as the pairwise distances are neglected. Moreover, the larger parameter $\alpha$ is, the larger aversion for long edges is. Since this family of functions (for various values of the parameter $\alpha$) captures a wide spectrum of behaviours, it should be enough to concentrate on this choice but one can easily experiment with other functions. So, for now we may assume that the only parameter of the model is $\alpha \in [0,\infty)$.

The \emph{Geometric Chung-Lu} (GCL) model is the random graph $G(\textbf{w}, \emb, \alpha)$ on the vertex set $V = \{ v_1, \ldots, v_n \}$ in which each pair of vertices $v_i, v_j$, independently of other pairs, forms an edge with probability $p_{i,j}$, where
\begin{equation*}
p_{i,j} = x_i x_j g(d_{i,j}) 
\end{equation*}
for some carefully tuned weights $x_i \in \R_+$. The weights are selected such that the expected degree of $v_i$ is $w_i$; that is, for all $i \in [n]$
$$
w_i = \sum_{j \in [n]} p_{i,j} =  x_i \sum_{j \in [n]} x_j g(d_{i,j}).
$$
Indeed, we prove in Appendix~\ref{apdx:model} that there exists the unique selection of weights, provided that the maximum degree of $G$ is less than the sum of degrees of all other vertices. Since each connected component of $G$ can be embedded independently, we may assume that $G$ is connected and so the minimum degree of $G$ is at least 1. As a result, this very mild condition is trivially satisfied unless $G$ is a star on $n$ vertices. Finally, let us mention that in Appendix~\ref{apdx:model} it is assumed that $g(d_{i,j}) > 0$ for all pairs $i,j$. In our case, $g(d_{i,j}) = 0$ for a pair of vertices that are at the maximum distance. It causes no problems in practice but, as mentioned earlier, one can easily scale the outcome of function $g(\cdot)$ to move away from zero without affecting the divergence score in any non-negligible way.


It is not clear how to find weights explicitly but they can be easily (and efficiently) approximated numerically to any desired precision.
In Appendix~\ref{apdx:model}, we prove that, if the solution exists, which is easy to check, then the set of right hand sides of the equations considered as a function from $\mathbb{R}^n$ to $\mathbb{R}^n$ is a local diffeomorphism everywhere in its domain. As a result, standard gradient root-finding algorithms should be quite effective in finding the desired weights. We discuss one simple numerical approximation procedure the next subsection.

\subsection{Numerical Approximation}

Let us start with an arbitrary vector $\textbf{t}^0=(t^0_1, \ldots, t^0_n)$ (say, $\textbf{t}^0=(1,\ldots,1)$) that we will carefully tune by iteratively constructing a sequence $(\textbf{t}^s)_{s \ge 0}$. Suppose that we are given a vector $\textbf{t}^s=(t^s_1, \ldots, t^s_n)$. If we introduce an edge between $v_i$ and $v_j$ with probability
\begin{equation}
p^s_{i,j} = t^s_i t^s_j g(d_{i,j}),
\label{eq:p2}
\end{equation}
then the expected degree of $v_i$ would be
\begin{equation}
s^s_i = \sum_j p^s_{i,j} = t^s_i \sum_j t^s_j g(d_{i,j}).
\label{eq:s}
\end{equation}
Note that, for a given vertex $v_i$, it easy to adjust the weights so that $s^s_i$ matches $w_i$, the desired expected degree of $v_i$; indeed, one can replace $t^s_i$ with $t^s_i (w_i/s^s_i)$. However, unfortunately, it will also affect other values of $\textbf{s}^s$ and vice versa, changes in other parts of $\textbf{t}$ affect $s^s_i$ too. Hence, instead of doing it, each vertex should take a small step into the right direction and this process should quickly converge to the desired state: $s^s_i$ being very close to $w_i$ for all $i$. Let us note that, for our application, we do not need to get close to the desired expected degree sequence. The divergence score we defined earlier is not too sensitive.
Fix some small constants $\eps, \delta > 0$.
For example, in our experiments we used $\eps = 0.1$ and $\delta = 0.001$.
For each $i$, $1 \le i \le n$, we define
\begin{equation}
t^{s+1}_i = (1-\eps) t^s_i + \eps t^s_i (w_i/s^s_i) = t^s_i + \eps t^s_i (w_i/s^s_i-1).
\label{eq:t}
\end{equation}
We repeat the tuning process until $\max_i |\textbf{w}_i - s^s_i| < \delta$.

\section{Experiments}\label{sec:experiments}

\subsection{Datasets}\label{sec:exp_data}

In order to test our algorithm, we benchmark it against synthetic graphs with communities as well as some real-world graphs with known community structures. This is a natural and often used approach (see, for example,~\cite{Lu2018}).
In all examples, we colour the vertices with respect to the known, ground-truth communities, but in the algorithm, we use the partition obtained via a graph clustering algorithm, as the algorithm is unsupervised.

\subsubsection{LFR}

As common in the literature, we analyze how the algorithm performs on artificially constructed networks with communities. 
A popular example is the LFR benchmark by Lancichinetti, Fortunato, and Radicchi~\cite{LFR} that generates synthetic graphs that resemble real world graphs. In this benchmark, the size of each community is drawn from a power-law distribution, as is the degree of each vertex. 

The LFR model has a number of parameters. The most important one is the mixing parameter $\mu$ that controls the fraction of {\it external} edges (that is, edges between communities).  Essentially this can be viewed as the amount of noise in the graph. In one extreme case, if $\mu=0$, then all the edges are within communities. On the other hand, if $\mu=1$, then all edges are between different communities. Other parameters control the number of vertices ($n$), the negative exponent of the power-law degree distribution ($\gamma_1$), the negative exponent of the distribution of community sizes ($\gamma_2$), and the average ($d$) and maximum degree ($d_{\max}$). 

We generated small LFR graphs with $n=100$ for visualization purpose.
For the mixing parameter $\mu$, we used $\mu=0.15$ (strong communities, low noise), $\mu=0.35$ (intermediate case), and $\mu=0.55$ (noisy graph, weaker communities).
For the other parameters, we used, $\gamma_1=2$, $\gamma_2=1$, $d=8$, and $d_{\max}=20$.   

\subsubsection{Zachary's Karate Club}

This well-known social network consists of 34 members of a karate club and 78 pairwise links observed over a period of three years. During the famous study of Zachary~\cite{Zachary77}, a political conflict arose between the club president and the instructor caused the club to split into two parts (the communities), 
each with half of the members. Zachary recorded a network of friendships among members of the club shortly before the fission.

\subsubsection{College Football}

This graph represents the schedule of United States football games between Division IA colleges during the regular season in Fall 2000~\cite{Football}. This is another well-studied, more complex, real-world network with known community structures.
The data consists of 115 teams (vertices) and 613 games (edges). 
 The teams are divided into conferences containing around 8–12 teams each. In general, games are more frequent between members of the same conference than between members of different conferences, with teams playing an average of about seven intra-conference games and four inter-conference games in the 2000 season. There are a few exceptions to this rule, as detailed in \cite{Lu2018}:
 one of the conferences is really a group of independent teams, one conference is really broken into two groups, and 3 other teams play mainly against teams from other conferences. We refer to those as {\it outlying} vertices, which we represent with a distinctive triangular shape.
 
\subsubsection{Email-Eu-core Network}

This network was generated using email data from a large European research institution and is available as one of the SNAP Datasets~\cite{snapnets}. There is a directed edge $(u, v)$ in the network if person $u$ sent person $v$ at least one email. The e-mails only represent communication between institution members (the core), and the dataset does not contain incoming messages from or outgoing messages to the rest of the world. The dataset also contains community memberships of the vertices: each individual belongs to exactly one of 42 departments at the research institute. 

There are 1005 vertices and 25571 edges in the original directed graph. After making it undirected and keeping only the largest connected component, we ended up with a graph on 986 vertices and 16064 edges. Most of the communities are very weak; in fact, only 1 community qualifies as a {\it weak community} as defined in equation (9.2) in \cite{Barabasi2016}. Those are communities for which the ratio between the total internal degree and the total degree is above 0.5. 
In our study, we looked at the communities for which this ratio was the highest.

\subsection{The Graph Embeddings}\label{sec:exp_emb}

We considered three popular graph embedding algorithms in our examples: 
\emph{node2vec}, VERSE and LINE. 
Recall that our goal is to experiment with our framework and not to do 
extensive comparison of graph embeddings. 
Comparisons are done in the original dimension of
the embedding. For visualization purposes, we project the embeddings into 2-dimensional space. 

\subsubsection{node2vec}

\emph{node2vec}~\cite{node2vec} is a semi-supervised algorithm for scalable feature learning in networks. Intuitively, the goal is to find feature representations that maximize the likelihood of preserving network neighbourhoods of vertices in a $d$-dimensional feature space. By choosing an appropriate notion of a neighborhood, node2vec can learn representations that organize vertices based on their network roles and/or communities they belong to. It is achieved by developing a family of biased random walks, which efficiently explore diverse neighborhoods of a given
vertex.

There are two main parameters in this algorithm. The ``return parameter'' $p$ controls the likelihood of immediately revisiting a vertex in the random walk.  Setting it to a high value ensures that we are less likely to sample an already-visited  vertex in the following two steps (unless, of course, the next vertex in the walk had no other neighbour). The ``in-out parameter'' $q$ allows the search to differentiate between ``inward'' and ``outward'' vertices. Loosely speaking, this parameter guides the neighborhood sampling strategy which allows us to smoothly interpolate
between BFS and DFS.
For \emph{node2vec}, we fixed parameter $q=1$ and we varied parameter $0.5 \le p \le 2$.
We considered embedding dimensions $2 \le D \le 128$.
For the other parameters we used the default values.

\subsubsection{VERSE}

VERtex Similarity Embeddings (VERSE)~\cite{verse} is a simple, versatile, and memory-efficient method that derives graph embeddings explicitly calibrated to preserve the distributions of a selected vertex-to-vertex similarity measure. It is a general framework that learns \emph{any} similarity measures among vertices via training a simple, yet expressive, single-layer neural network. This includes popular similarity measures such as Personalized PageRank
(PPR), SimRank, and adjacency similarity. 
We used the default recommended parameters for this algorithm and
we considered embedding dimensions $2 \le D \le 128$.

\subsubsection{LINE}

Large-scale Information Network Embedding (LINE) \cite{LINE} is an efficient method for vertex embedding that uses an approximate factorization of the adjacency matrix, trying to preserve first as well as second order proximities.
We used the default recommended parameters for this algorithm, but we varied the number of threads used from 1 (default) to 10. We considered embedding dimensions $2 \le D \le 128$.

\subsection{Visualization}

Dimension reduction seeks to produce a low dimensional representation of high
dimensional data that preserves relevant structure. As a result, it is an important problem in data science as a potential pre-processing step for many machine learning algorithms. Here, we use it for visualization of tested graph embeddings that are done in high dimensions.
We used UMAP (Uniform Manifold Approximation and Projection)~\cite{UMAP}, a novel manifold learning technique for dimension reduction. UMAP is constructed from a theoretical framework based in Riemannian geometry and algebraic topology. It provides a practical scalable algorithm that applies to real world datasets.

\subsection{Results}

\subsubsection{Artificial Graphs}

\begin{figure}[ht]
\centering
\hbox{
\includegraphics[width=8cm]{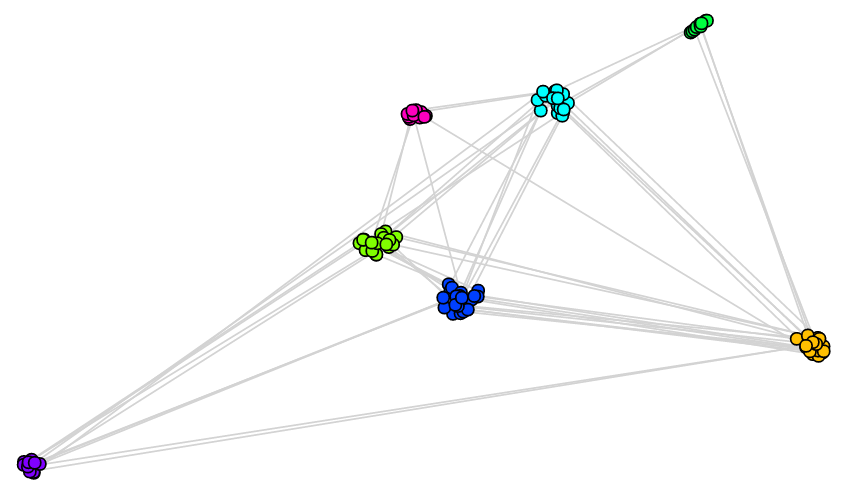}
\includegraphics[width=8cm]{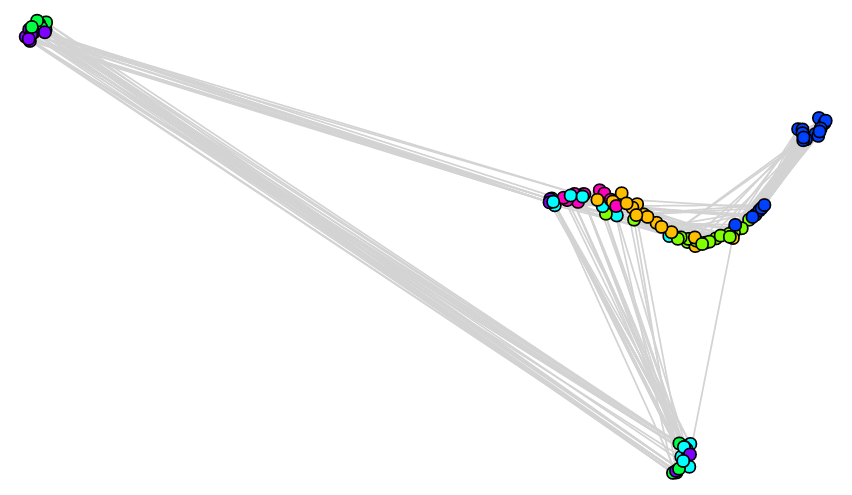}
}
\vspace{0.25cm}
\hbox{
\includegraphics[width=8cm]{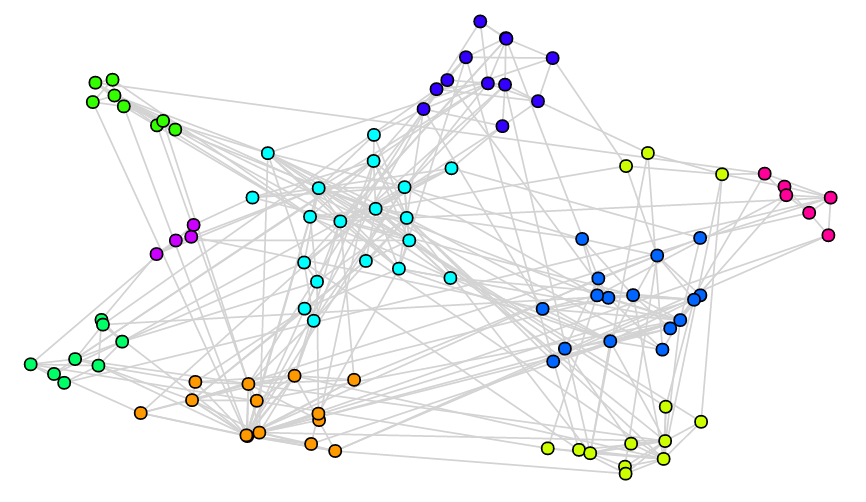}
\includegraphics[width=8cm]{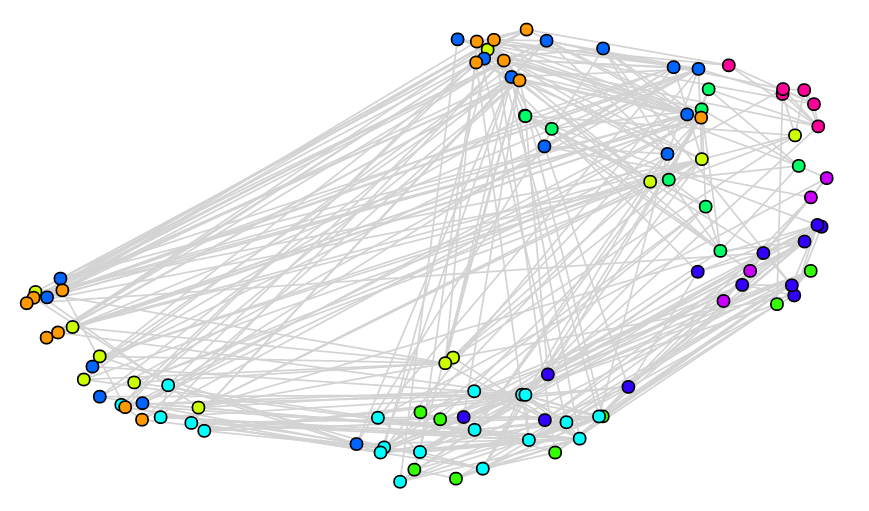}
}
\vspace{0.25cm}
\hbox{
\includegraphics[width=8cm]{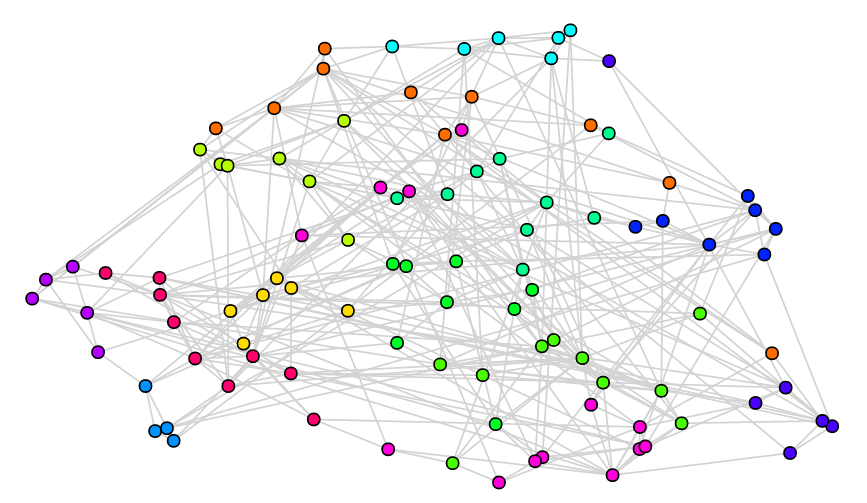}
\includegraphics[width=8cm]{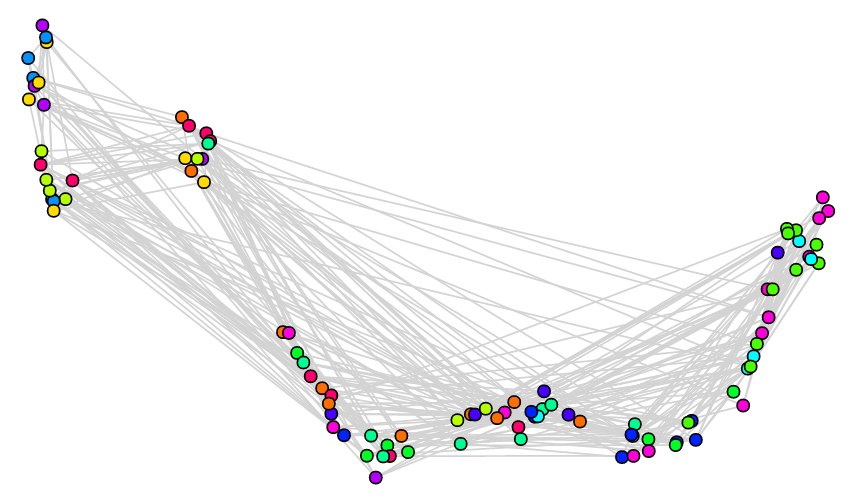}
}
\caption{Best (left column) and worst (right column) scoring embeddings for 3 different LFR graphs.
In the top row, we show the results for the graph with 7 strong communities ($\mu=0.15$).
In the middle row, we show the results for the graph with $\mu=0.35$ and 9 communities.
Finally in the bottom row, we show the results for the noisy graph ($\mu=0.55$) with 13 communities.
}
\label{fig:lfr}
\end{figure}

For the three LFR graphs described earlier, we generated 5 embeddings for every choice of parameter(s) and for each algorithm considered, for a total of 315 embeddings for each graph.
In Figure~\ref{fig:lfr}, we plot a 2-dimension 
projection of the best and worst results as identified by our divergence score.
The colors correspond to the ground-truth communities generated by the LFR benchmark.
For the best embeddings, we clearly see that the ground-truth communities are grouped together, even (but to a lesser extent) for the noisy graph with $\mu=0.55$. This is not the case for the worst embeddings.

\clearpage

\subsubsection{Zachary's Karate Club Network}

\begin{figure}[ht]
\centering
\hbox{
\includegraphics[width=8cm]{karate-best.png}
\includegraphics[width=8cm]{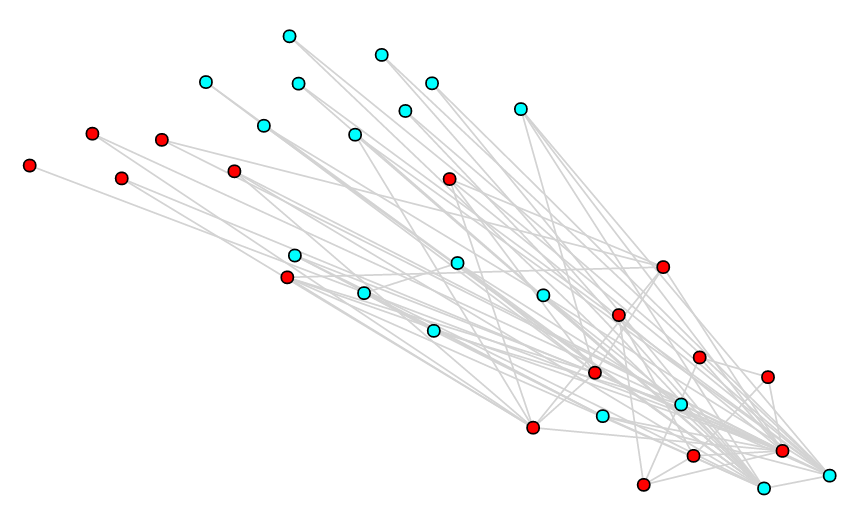}
}
\caption{Zachary's Karate Club. We show the best (left) and the worst (right) embedding according to our algorithm. The left plot is clearly a better choice.}
\label{fig:karate-best-worst}
\end{figure}

We already considered this dataset in Section \ref{sec:framework}.
In Figure \ref{fig:karate-best-worst}, we show the best and worst scoring embeddings we found using our algorithm. The colors correspond to the 2 real communities.

\subsubsection{The College Football Graph}

\begin{figure}[ht]
\centering
\hbox{
\includegraphics[width=8cm]{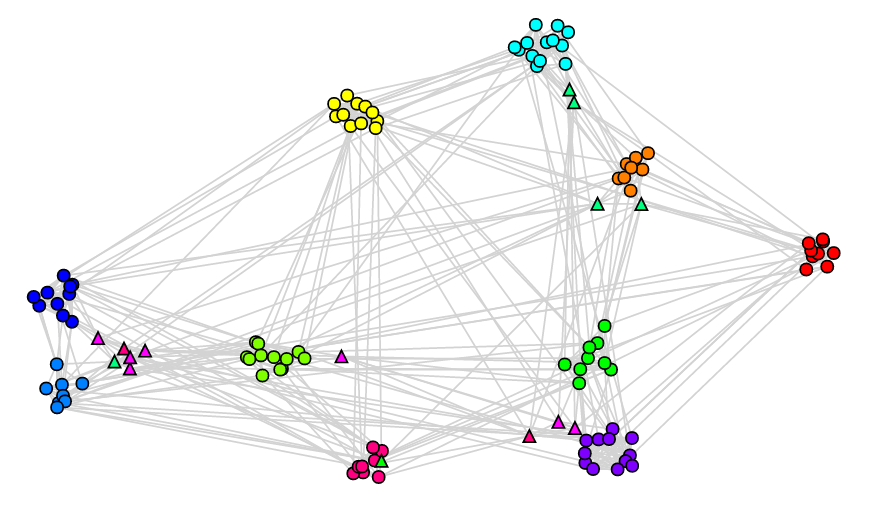}
\includegraphics[width=8cm]{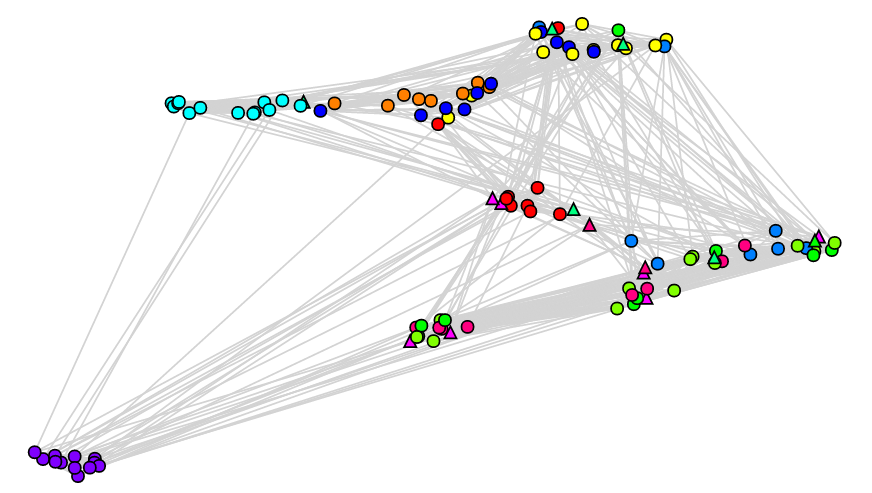}
}
\caption{The College Football Graph. We show the best scoring embedding (left) and the worst one (right). The communities are very clear in the left plot.}
\label{fig:foot_best}
\end{figure}

This graph consists of 115 teams spread between 12 conferences.
In Figure \ref{fig:foot_best}, we show the best and worst scoring embeddings. The colors of vertices correspond to the conferences, white  triangular shaped vertices correspond to outlying vertices as explained earlier. 

\begin{figure}[ht]
\centering
\hbox{
\includegraphics[width=8cm]{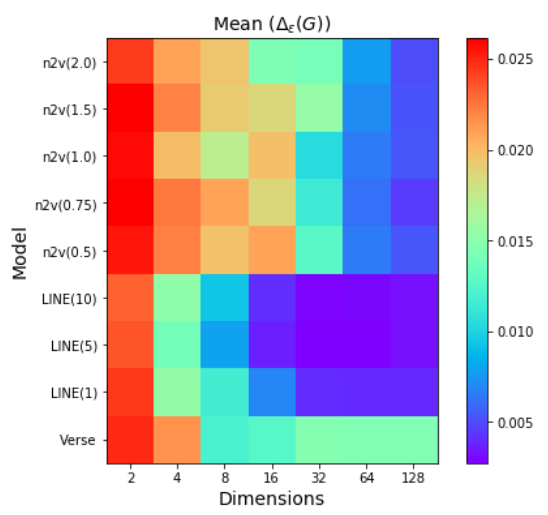}
\includegraphics[width=8cm]{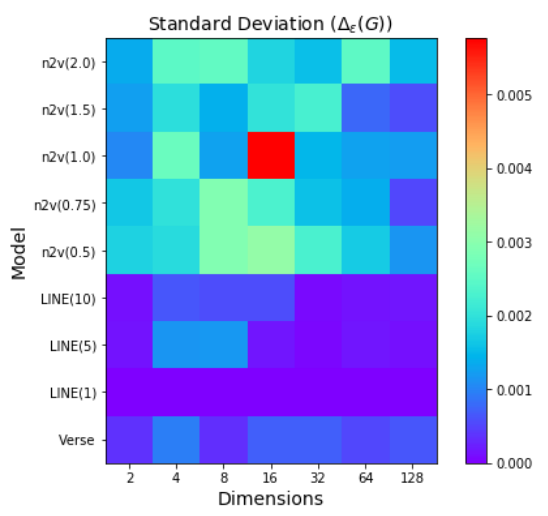}
}
\caption{
For the College Football Graph, for each choice of algorithm, parameters and dimension, we generated 5 embeddings. We show the mean and standard deviation of the divergence score $\Delta_\emb(G)$ for each of those.
}
\label{fig:foot_div}
\end{figure}

For each choice of embedding model, parameters and dimension, we obtained 5 different embeddings. In Figure \ref{fig:foot_div}, we summarize those results by showing the mean and standard deviation of the divergence score $\Delta_\emb(G)$.
This allows view some overall properties: for example, embedding in 2-dimensions is not enough in this case; we obtained the best result with LINE with 32 dimensions or more. Recall that we are not performing a overall study of graph embedding algorithms, and we are not looking for a winning model. In fact, both VERSE and {\it node2vec} gave the best results for other graphs, which indicates the need for a comparison framework.

\subsubsection{The email-Eu-core Graph}

\begin{figure}[ht]
\begin{center}
\hbox{
\includegraphics[width=14cm]{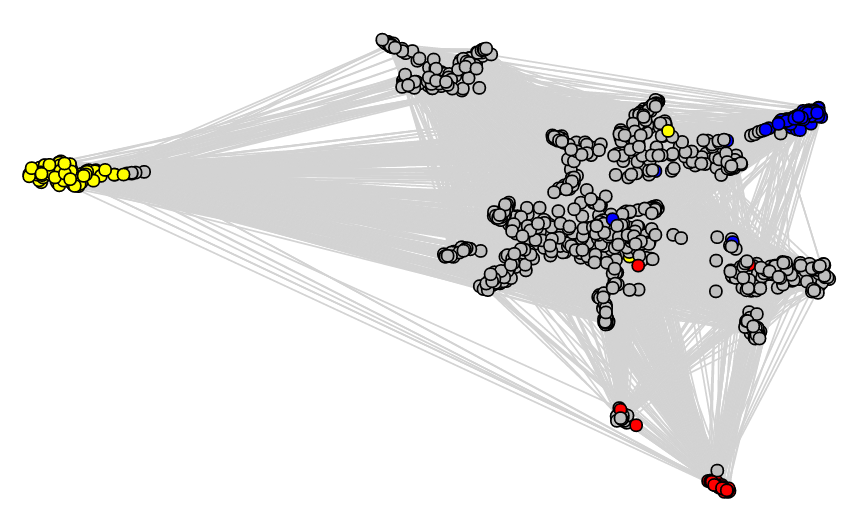}
}
\end{center}
\caption{The best embedding we obtained for the email-Eu-core graph with the 3 strongest communities respectively shown in yellow, red and blue, which are clearly separated.}
\label{fig:email}
\end{figure}

In Figure~\ref{fig:email}, we show the highest scoring embedding we obtained for the email graph described earlier.  
Since most communities are very weak,
we highlight only the 3 strongest communities for which the ratio of internal to total degree are respectively 0.54, 0.42 and 0.39.

\subsubsection{Graph Clustering}

An important part of our algorithm consists of building a reasonable partition of the vertices, over which we build the representative vectors~(\ref{eq:c}) and~(\ref{eq:b}). 
The ECG algorithm was shown 
to generally yield good and stable clusters~\cite{ECG}, and we used it throughout. 
We re-ran all embeddings on the College Football graph using respectively the Louvain~\cite{louvain} and InfoMap~\cite{infomap} clustering algorithms, and in each case, we produced a ranking of the embeddings with respect to our method. We compared those rankings as well as the ranking obtained with the ECG clustering algorithm by computing the Kendall-tau correlation between those. The results, which are summarized in Table~\ref{table:tau}, show very high correlation. 

\begin{table}[ht]
\begin{center}
\begin{tabular}{|*{3}{c|}}
  \cline{1-3}
  {\bf Algorithms} & {\bf ECG} & {\bf Louvain} \\ \cline{1-3}
  {\bf Louvain} & 0.81 & \\ \cline{1-3}
  {\bf InfoMap} & 0.83 & 0.79 \\ \cline{1-3}
\end{tabular}
\end{center}
\caption{Kendall-tau correlation between all ranked embeddings on the College Football graph using 3 different graph clustering algorithms.}
\label{table:tau}
\end{table}

\section{Future Directions}
\label{sec:conclusion}

In this paper, our aim was to introduce a general framework for evaluating embeddings. This exploratory research showed that our algorithm is very promising. The next natural step is to do extensive experiments of various embedding algorithms on large real-world datasets in order to evaluate and compare them.

Hypergraphs are natural generalizations of graphs in which a single (hyper)edge can connect any number of vertices. As a consequence, hypergraphs are often more suitable and useful to model many important networks and processes. Typical applications are related to social data analysis and include situations such as exchanging emails with several recipients, co-authorship of research papers, or analyzing security vulnerabilities of information networks. In many situations, using hypergraphs instead of classical graphs allows us to better capture and analyze dependencies within the network. We are interested in generalizing classic notions to hypergraphs, such as clustering via modularity~\cite{Modularity}, as well as developing new algorithms for them~\cite{Hypergraphs}. Hence, a natural line of development of the proposed embedding algorithm is to generalize it to allow for evaluation of embeddings of hypergraphs.

As a side effect of our research on graph embeddings, we have introduced the Geometric Chung-Lu model that is interesting on its own rights and potentially applicable in other problems. As it is not the main focus of this paper, we did not analyze its graph-theoretic properties in detail. Their study remains as a subject for further research. The properties, that are standard in the analysis of other models of complex networks, include clustering coefficient, connectivity, average path length, centrality measures.

\newpage

\appendix

\section{Appendix --- Geometric Chung-Lu Model is Well-defined} \label{apdx:model}

Let us state the problem in a slightly more general situation. Suppose that for each pair $i, j \in [n]$ we have $a_{ij}=a_{ji} \in \R_+$, and for each $i \in [n]$ we have $a_{ii}=0$ and $b_i \in \R_+$. In our application, elements of vector $\mathbf{b} = (b_i)_{i\in[n]}$ satisfy $b_i = w_i \in [n-1]$ and correspond to the degree distribution, and elements of matrix $A$ satisfy $a_{i,j} = g(d_{i,j}) \in (0,1]$ and correspond to the embedding distance between vertices. Our goal is to investigate if there is a  solution, $x_i \in \R_+$ for $i \in [n]$, of the following system of equations:
\begin{equation}\label{eq:system}
b_i = x_i \sum_{j=1}^na_{ij}x_j \quad \text{ for all } i\in[n].
\end{equation}
If there is one, then is this solution unique?

\medskip

The $n=2$ case a degenerate case that exhibit a different behaviour but is easy to investigate: $b_1 = x_1 x_2 a_{12}$ and $b_2 = x_2 x_1 a_{21}$. It follows that the solution exists if and only if $b_1 = b_2$. If this is the case, then there are infinite number of solutions, each of them of the form $(x_1, x_2) = (t,b_1/(a_{12}t))$ for some $t \in \R_+$.

Suppose now that $n \ge 3$. We will show that the desired solution of~(\ref{eq:system}) exists if and only if
\begin{equation}\label{eq:condition}
\sum_{i=1}^nb_i > 2\max_{i\in[n]}b_i.
\end{equation}
In other words, the condition is that the maximum element in vector $\mathbf{b}$ is smaller than the sum of the other elements. This is a very mild condition that holds in our application.  More importantly, this solution is unique. 

\medskip

We will start with proving uniqueness. After that, we will show that~(\ref{eq:condition}) is indeed an obvious, necessary condition before proving that it is also a sufficient one.  

\subsection{Uniqueness}

Let us assume that $n \ge 3$. For a contradiction, suppose that we have two different solutions: $\mathbf{x} = (x_i)_{i\in[n]}$ ($x_i \in \R_+$, $i \in [n]$) and $\mathbf{y} = (y_i)_{i \in [n]}$ ($y_i \in \R_+$, $i \in [n]$). It follows that for all $i \in [n]$ we have
$$
b_i = f_i(\mathbf{x}) = f_i(\mathbf{y}), \quad \text{ where } f_i(\mathbf{x}) = x_i \sum_{j=1}^n a_{ij} x_j.
$$
Let us analyze what happens at point $\mathbf{z} = t\mathbf{x} + (1-t)\mathbf{y}$ for some $t \in [0,1]$ (that is, $z_i = tx_i+(1-t)y_i$, $i \in [n]$). For each $i \in [n]$ we get
\begin{eqnarray*}
f_i(\mathbf{z}) &=& (tx_i+(1-t)y_i) \sum_{j=1}^n a_{ij} (tx_j+(1-t)y_j) \\
&=& \sum_{j=1}^n a_{ij} \left( t^2 x_i x_j+ t(1-t)(x_i y_j + x_j y_i) + (1-t)^2 y_i y_j \right) \\
&=& f(\textbf{x}) t^2 + \frac {t(1-t)}{x_i y_i} (f(\textbf{y}) x_i^2 + f(\textbf{y}) y_i^2) + f(\textbf{y}) (1-t)^2 \\
&=& b_i \left( t^2 + \frac {t(1-t)}{x_i y_i} (x_i^2 + y_i^2) + (1-t)^2 \right) \\
&=& b_i \left( 1 - 2t(1-t) + \frac {t(1-t)}{x_i y_i} (x_i^2 + y_i^2) \right) \\
&=& b_i \left( 1 + \frac {t(1-t)}{x_i y_i} (x_i^2 - 2x_iy_i + y_i^2) \right) \\
&=& b_i \left( 1+t(1-t)\frac{(x_i-y_i)^2}{x_iy_i} \right) =: h_i(t).
\end{eqnarray*}
Note that $h_i'(1/2)=0$ for all $i$ (as either $x_i-y_i$ vanishes and so $h_i(t)$ is a constant function or it does not vanish but then $h_i(t)$ is a parabola with a maximum at $t=1/2$). For convenience, let $\mathbf{v} = (\mathbf{x} + \mathbf{y})/2$ and $\mathbf{s} = (\mathbf{x} - \mathbf{y})/2$ (that is, $v_i=(x_i+y_i)/2$ and $s_i=(x_i-y_i)/2$ for all $i \in [n]$). It follows that
$$
\frac{df_i}{dh} \Big( \mathbf{v}+h\mathbf{s} ~|~ h=0 \Big)=0.
$$
On the other hand,
$$
f_i(\mathbf{v}+h\mathbf{s}) = \sum_{j=1}^na_{ij}(v_i+hs_i)(v_j+hs_j)
$$
so
$$
\frac{df_i}{dh}(\mathbf{v}+h\mathbf{s})=
\sum_{j=1}^na_{ij}(s_j(v_i+hs_i)+s_i(v_j+hs_j)).
$$
Combining the two observations, we get that
$$
0=\frac{df_i}{dh} \Big( \mathbf{v}+h\mathbf{s} ~|~ h=0 \Big) = 
\sum_{j=1}^na_{ij}(s_jv_i+s_iv_j)=
s_i\sum_{j=1}^na_{ij}v_j+\sum_{j=1}^ns_ja_{ij}v_i.
$$
Since $a_{ii}=0$, it is equivalent to 
$$
\mathbf{g}_i \ \mathbf{s}= 0,
$$
where $\mathbf{g}_i$ is a row vector such that $g_{ij}=a_{ij}v_i$ if $i\neq j$ and $g_{ii}=\sum_{j=1}^na_{ij}v_j$. Since this is true for all $i \in [n]$, we get
$$
G\mathbf{s}=\mathbf{0},
$$
where $g_{ij}$'s, the elements of matrix $G$, are defined as above. Since our assumption is that $\mathbf{x} \neq \mathbf{y}$, we have that $\Vert s \Vert >0$ and so we get that $\det(G)=0$.

Now, as multiplying a column of a matrix by a positive constant does not change the sign of its determinant, we may multiply column $j$ by $v_j > 0$ to conclude that
$$
\det(C)=0,
$$
where $c_{ii}=\sum_{j=1}^na_{ij}v_iv_j$ and $c_{ij}=a_{ij}v_iv_j$ for $i\neq j$. Thus, $C$ is a symmetric matrix whose diagonal is a sum of all its entries in a row off diagonal.

Now, since $\det(C)=0$, we get that there exists a non-zero vector $\mathbf{u}$ such that $C\mathbf{u}=\mathbf{0}$. Without loss of generality, we may assume that the first entry of this vector is a largest one in absolute value. Since we may scale this vector, without loss of generality, we may also assume that $\Vert \mathbf{u} \Vert_\infty=1$; that is, $u_1=1$ and $|u_i|\leq1$ for all $i$. 
Let us consider the product of the $1$st row of $C$ and vector $\mathbf{u}$. We have that
$$
0 = 1\cdot\sum_{j=2}^na_{1j}v_1v_j + \sum_{j=2}^n u_ja_{1j}v_1v_j 
= \sum_{j=2}^n (1+u_j)a_{1j}v_1v_j.
$$
But this means that $u_j=-1$ for $j>1$ (since all other numbers involved are positive). Finally, by considering the $2$nd row of the product $C\mathbf{u}=\mathbf{0}$, we get that 
$$
0 = a_{21}v_2v_1 - \sum_{j \neq 2} a_{2j}v_2v_j - \sum_{j=3}^n a_{2j}v_2v_j = -2 \sum_{j=3}^n a_{2j}v_2v_j.
$$
This is possible only if $n=2$; that is, when the sum is trivial. Since $n\ge 3$, we get the desired contradiction. This finishes the proof of the uniqueness of the solution of~(\ref{eq:system}) for $n \ge 3$.

Also note that $G$ is the Jacobian matrix of the function $[f_i(\mathbf{x})]$ and from the proof it follows that $\det(G)\neq 0$ in all admissible points where the system of equations $b_i=f_i(\mathbf{x})$ has a solution.

\subsection{Necessity}

Without loss of generality, we may assume that the sequence of $b_i$'s is non-increasing. In particular, $b_1$ is a largest value. Suppose that $x_i \in \R_+$, $i \in [n]$, is a solution of the system~(\ref{eq:system}). Since $a_{ij}=a_{ji}>0$ for $i\neq j$ and $a_{ii}=0$, we observe that for $n\geq 3$
$$
\sum_{i=2}^n b_i 
= \sum_{i=2}^n \left( x_i \sum_{j=1}^na_{ij}x_j \right)
> \sum_{i=2}^n \left( x_i a_{i1} x_1 \right)
= x_1 \sum_{i=1}^n a_{1i} x_i = b_1.
$$
Hence, we get that~(\ref{eq:condition}) is a necessary condition for the existence of the solution.

\subsection{Sufficiency}

In this subsection, we will still assume that $n\geq 3$. For a contradiction, suppose that there exists a vector $\mathbf{b} = (b_i)_{i \in [n]}$, with $b_i>0$ for all $i$, that satisfies~(\ref{eq:condition}) but for which there is no solution to the system~(\ref{eq:system}). We will call such vectors \emph{infeasible}; on the other hand, vectors that yield a solution $\mathbf{x} = (x_i)_{i \in [n]}$, with $x_i>0$ for all $i$, will be called \emph{feasible} (as proved above, the solution must be unique). Without loss of generality, we may assume that $b_1$ is a largest value in vector $\mathbf{b}$. 

Let us now construct another vector $\mathbf{b}' = (b'_i)_{i \in [n]}$ for which there is a solution to~(\ref{eq:system}) (that is, $\mathbf{b}'$ is feasible) but also $b'_1=b_1$ is a largest element in $\mathbf{b}'$. Indeed, it can be done easily by, for example, taking $x'_1=s$ for some large enough $s \in \R_+$ and $x'_2 = \ldots = x'_n = b_1 / (s \sum_{j \in [n]} a_{1j})$. We immediately get that
$$
b'_1 = x'_1 \sum_{j \in [n]} a_{1j} x'_j = b_1
$$
and for $i \ge 2$ we have
$$
b'_i = x'_i \sum_{j \in [n]} a_{ij} x'_j = x'_i a_{i1}x'_1 + x'_i \sum_{j=2}^n a_{ij} x'_j = b_1\frac {a_{1i}}{\sum_{j \in [n]} a_{1j}} + \left( \frac {b_1}{s \sum_{j \in [n]} a_{1j}}\right)^2 \sum_{j=2}^n a_{ij} < b_1,
$$
provided that $s \in \R_+$ is sufficiently large (since $n\ge 3$, we get that $a_{1i}/\sum_{j \in [n]} a_{1j} < 1$, and the second term above tends to zero as $s \to \infty$).

We will consider points along the line segment between $\mathbf{b}'$ and $\mathbf{b}$, namely, 
$$
\mathbf{b}(t) = (b_i(t))_{i \in [n]} = (1-t)\mathbf{b}'+t\mathbf{b}, \qquad \text{ for } t\in[0,1].
$$
Since $\mathbf{b}'$ is feasible and we already proved that~(\ref{eq:condition}) is a necessary condition, $\mathbf{b}'$ satisfies~(\ref{eq:condition}). But, not only $\mathbf{b}$ and $\mathbf{b}'$ satisfy it but also $\mathbf{b}(t)$ satisfies it for any $t \in [0,1]$. Let us fix any $t \in (0,1)$. Clearly, $b_1(t)=b_1=b'_1$ is a largest value in $\mathbf{b}(t)$. More importantly,
$$
\sum_{i \in [n]} b_i(t) = (1-t) \sum_{i \in [n]} b'_i + t \sum_{i \in [n]} b_i > (1-t)\cdot 2 \max_{i \in [n]} b'_i + t \cdot 2 \max_{i \in [n]} b_i = 2 b_1(t) = 2 \max_{i\in [n]} b_i(t)
$$
and so, indeed,~(\ref{eq:condition}) holds for $\mathbf{b}(t)$. It follows that there exists a universal $\eps > 0$ such that for any $t \in [0,1]$ we have
\begin{equation}\label{eq:epsilon}
2 \max_{i \in [n]} b_i(t) < (1-\eps) \sum_{i \in [n]} b_i(i).
\end{equation}

Fix $t \in (0,1)$ and suppose that $\mathbf{b}(t)$ is feasible. Let $\mathbf{x}(t) = (x_i(t))_{i\in [n]}$ be the (unique) solution for $\mathbf{b}(t)$. Let $G(t) = \left( \frac {\partial f_i}{\partial x_j} \right)_{ij}$ be the Jacobian matrix of our transformation function at point $\mathbf{x}(t)$; that is, $g_{ij}(t) = a_{ij} x_i(t)$ if $i \neq j$ and $g_{ii}(t) = \sum_{j \in [n]} a_{ij} x_j(t)$. From the analysis performed in the proof of uniqueness of the solution it follows that $\det(G(t))\neq 0$ and so our transformation is a local diffeomorphism. As a result, any open set in $\R^n$ containing $\mathbf{x}(t)$ is mapped to an open set in $\R^n$ containing $\mathbf{b}(t)$. In particular, there exists $\delta > 0$ such that $\mathbf{b}(s)$ is feasible for any $t - \delta \le s \le t + \delta$. Combining this observation with the fact that $\mathbf{b}$ is \emph{not} feasible we get that there exists $T \in (0,1]$ such that $\mathbf{b}(T)$ is not feasible but $\mathbf{b}(t)$ is feasible for any $t \in [0,T)$.

Consider then any sequence $(t_i)_{i \in \N}$ of real numbers $t_i \in [0,T)$ such that $t_i\to T$ as $i \to \infty$; for example, $t_i = T(1-1/i)$. All limits from now on will be for $i \to \infty$. Recall that $\mathbf{b}(t_i)$ is feasible and so $\mathbf{x}(t_i)$ is well-defined.

Let us first note that it is impossible that $\Vert \mathbf{x}(t_i) \Vert$ is bounded for infinitely many $i$. Indeed, if it were the case, then by Bolzano-Weierstrass theorem it would have a subsequence $(\mathbf{x}(t_{s_i}))_{i \in [n]}$ such that $\Vert \mathbf{x}(t_{s_i}) \Vert \to c \in \R$. However, then, by continuity of our transformation, the limit value $\mathbf{b}(T)$ would be feasible. Note, in particular, that it is impossible that $x_j(t_{s_i}) \to 0$ for some $j$ as then the corresponding $b_j(t_{s_i})$ would also tend to zero which implies that $b_j(T) = 0$; this is not possible as both $b_j$ and $b'_j$ are bounded away from zero.

It follows that $\Vert \mathbf{x}(t_i) \Vert \to \infty$. This means that there exist $p \in [n]$ and subsequence $(t_{s_i})_{i \in \N}$ such that $x_p(t_{s_i})\to \infty$. Let us now observe that it is not possible that $x_p(t_{s_i})\to \infty$ and $x_q(t_{s_i})$ does not tend to zero for some $q\neq p$. Indeed, since $b_p(t_{s_i}) \ge a_{pq} x_q(t_{s_i}) x_p(t_{s_i})$, this would imply that for any constant $c$, $b_p(t_{s_i}) > c$ for an infinite number of $i$'s. This is impossible as $b_p(t_{s_i}) \to b_p(T) \le b_1(T)$. We get that $x_p(t_{s_i})\to \infty$ for precisely one index $p$ and $x_q(t_{s_i})\to 0$ for all $q\neq p$.

Now, for a given $q \neq p$, note that
$$
\sum_{j\neq p} a_{qj} x_q(t_{s_i}) x_j(t_{s_i}) \to 0
$$
and
$$
a_{qp} x_q(t_{s_i}) x_p(t_{s_i}) + \sum_{j\neq p} a_{qj} x_q(t_{s_i}) x_j(t_{s_i}) = \sum_{j \in [n]} a_{qj} x_q(t_{s_i}) x_j(t_{s_i}) = b_q(t_{s_i}) \to b_q(T).
$$
We get that for sufficiently large values of $i$, $a_{qp} x_q(t_{s_i}) x_p(t_{s_i}) > (1-\eps/2) b_q(T)$, where $\eps > 0$ is the same as in~(\ref{eq:epsilon}). It follows that for all sufficiently large $i$,
$$
b_p(t_{s_i}) = \sum_{q \neq p} a_{pq} x_p(t_{s_i}) x_q(t_{s_i}) \ge \left( 1 - \frac {\eps}{2} \right) \sum_{q \neq p} b_q(T)
$$
and so also in the limit
$$
b_p(T) \ge \left( 1 - \frac {\eps}{2} \right) \sum_{q \neq p} b_q(T).
$$
But this means that
$$
2 \max_{i \in [n]} b_i(T) \ge 2 b_p(T) \ge b_p(T) + \left( 1 - \frac {\eps}{2} \right) \sum_{q \neq p} b_q(T) \ge \left( 1 - \frac {\eps}{2} \right) \sum_{q \in [n]} b_q(T)
$$
that contradicts~(\ref{eq:epsilon}). Hence, we get that~(\ref{eq:condition}) is also a sufficient condition for the existence of the solution.

\section{More Illustrations of the GCL Model} 
\label{apdx:plots}

In Section \ref{sec:framework}, we illustrated our framework with the
small Zachary's Karate Club graph.
In this section, we provide the same illustration for all the other
graphs we considered except for the larger email graph which is less amenable
to nice visualization.

For each graph considered, we select the best scoring embedding
we found, and we plot the divergence score as a function of the 
$\alpha$-parameter that we use in our framework. Recall that we search
over a grid of values for $\alpha$ and we select the one for which the 
divergence score in (\ref{eq:delta}) is minimized. We also plot the graph
itself using this best embedding.

The Geometric Chung-Lu model computes probabilities of seeing edges between
vertices as a function of the degrees of the corresponding vertices and their distance in embedded
space, the latter being a function of $\alpha$. 
Thus, we can use this model to generate random graphs given the degrees of vertices and
some embedding.
With small values of $\alpha$,
the importance of the distances is small (at the extreme case, when $\alpha=0$, it is completely neglected), 
and several long edges (between communities) are generated.
With large values of $\alpha$, it is the opposite and we expect much
fewer edges between communities.
The optimal value, which we denote $\hat{\alpha}$, aims at preserving the right
balance between long and short edges (respectively between and within communities).

For each graph considered, we show three generated random graph respectively 
using $\alpha=0, \alpha = \hat{\alpha}$, and some value of $\alpha > \hat{\alpha}$.

\clearpage

\subsection{The College Football Graph}

\begin{figure}[ht]
\begin{center}
\hbox{
\includegraphics[width=7.5cm]{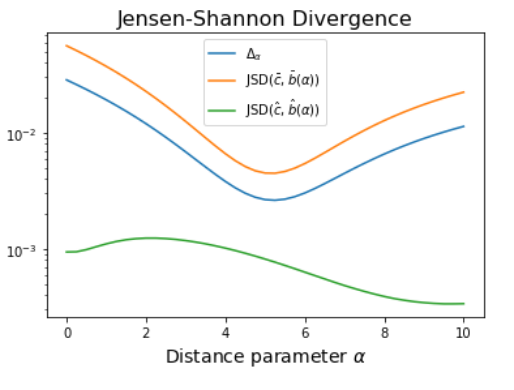}
\includegraphics[width=7.5cm]{football-best.png}
}
\vspace{1cm}
\hbox{
\includegraphics[width=5.4cm]{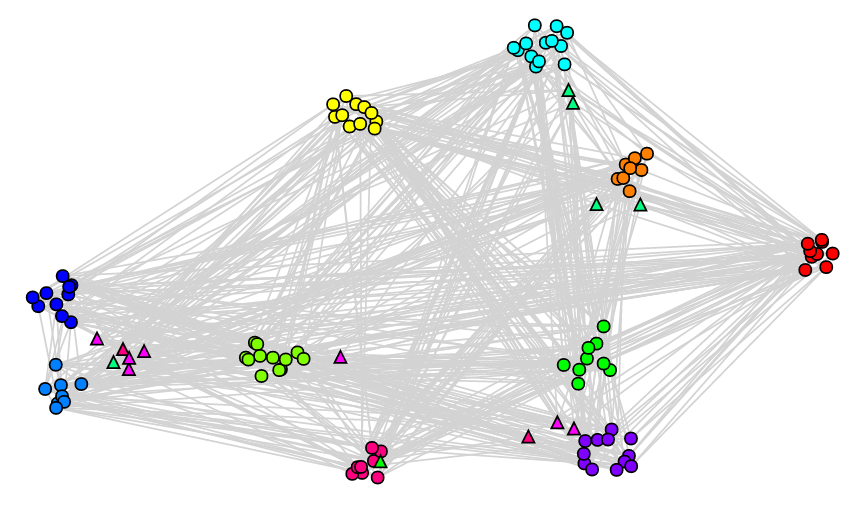}
\includegraphics[width=5.4cm]{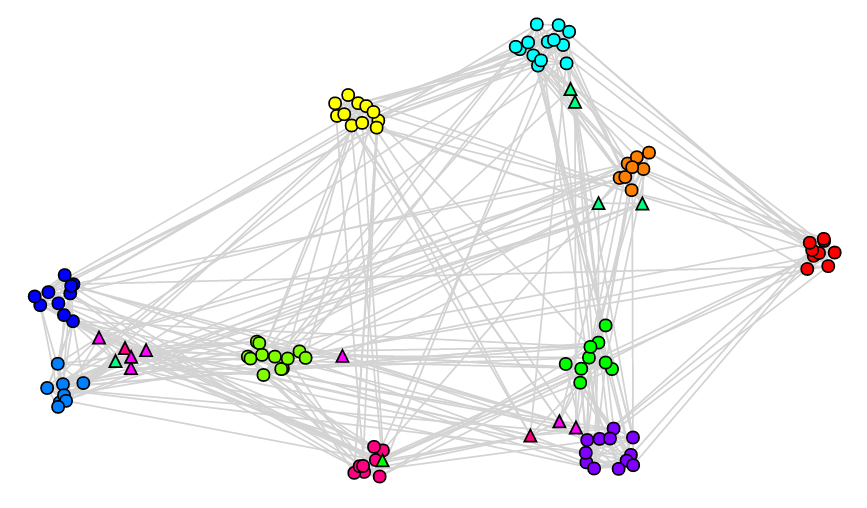}
\includegraphics[width=5.4cm]{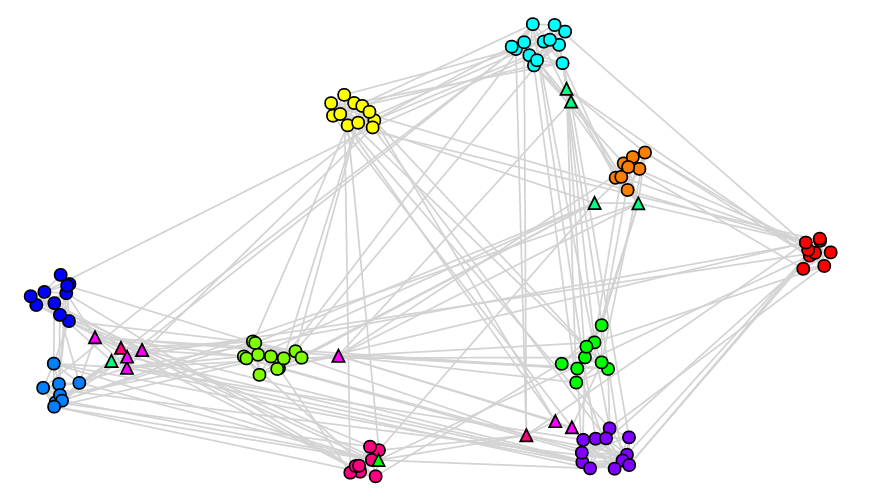}
}
\end{center}
\caption{The College Football Graph. On the top row, we illustrate the divergence score as a function of $\alpha$ (left) for the best embedding found by our framework (right). 
In the bottom row, we generate edges using the Geometric Chung-Lu Model respectively with $\alpha=0$ (left), the optimal value $\hat{\alpha}=5.25$ (center) and $\alpha=7$ (right).  
\label{fig:football-alpha}}
\end{figure}

For this graph, most communities are very strong, and the best embedding puts their vertices quite close, as we see in Figure \ref{fig:football-alpha}.
The impact of the choice of $\alpha$ for the distance function in embedded space is clearly seen in the bottom row of Figure \ref{fig:football-alpha}. 
With $\alpha=0$, there are many edges between communities, as only the expected degrees are preserved. For $\alpha$ larger than the optimal value, we see the opposite behaviour. The random graph generated with the optimal $\alpha$ value is visually very similar to the real graph shown in the top row.

\clearpage


\subsection{LFR Graph with Strong Communities}

\begin{figure}[ht]
\begin{center}
\hbox{
\includegraphics[width=7.5cm]{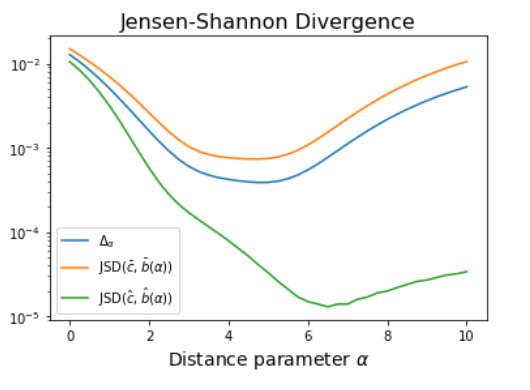}
\includegraphics[width=7.5cm]{LFR15a-best.png}
}
\vspace{1cm}
\hbox{
\includegraphics[width=5.4cm]{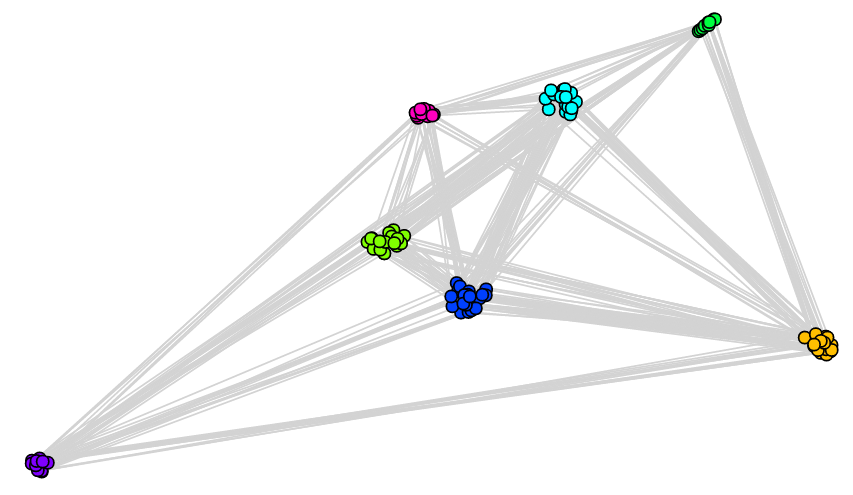}
\includegraphics[width=5.4cm]{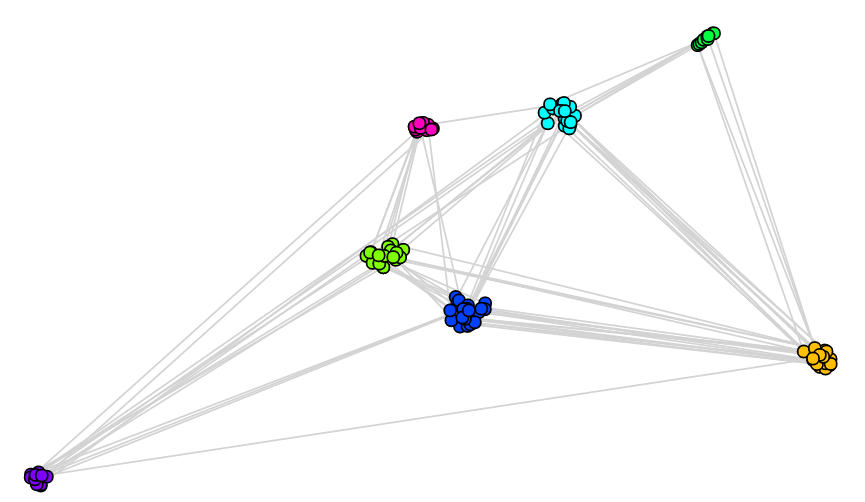}
\includegraphics[width=5.4cm]{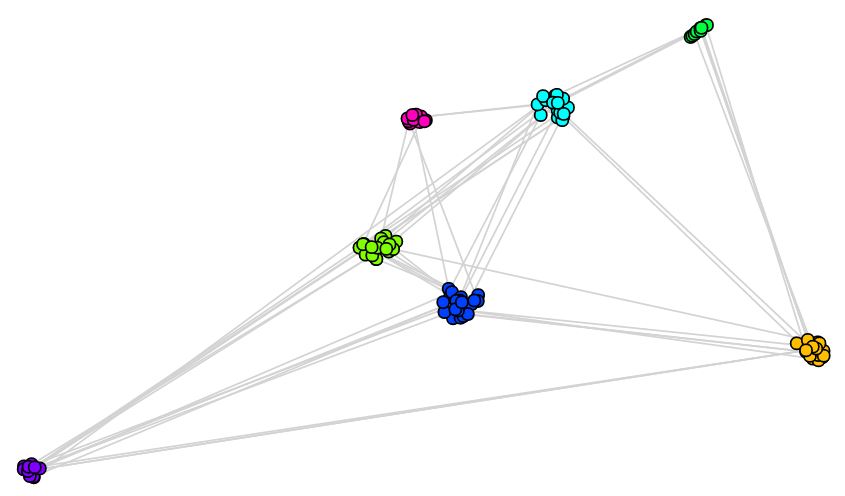}
}
\end{center}
\caption{For the LFR graph with mixing parameter $\mu=0.15$, on the top row, we illustrate the divergence score as a function of $\alpha$ (left) for the best embedding found by our framework (right).
On the bottom row, we generate edges via the
Geometric Chung-Lu Model with distance parameters $\alpha=0$ (left),
the optimal value $\hat{\alpha}=4.75$ (center) and $\alpha=7$ (right).
\label{fig:lfr15a-alpha}}
\end{figure}

The communities are very strong given the choice of value $\mu=0.15$, so good embeddings should put their vertices quite close, as we see in Figure \ref{fig:lfr15a-alpha}.
The impact of the choice of $\alpha$ for the distance function in embedded space is clearly seen in the bottom row of Figure \ref{fig:lfr15a-alpha}, with the same conclusion as for the College Football graph. 

\clearpage


\subsection{LFR Graph with Weaker Communities}

\begin{figure}[ht]
\begin{center}
\hbox{
\includegraphics[width=7.5cm]{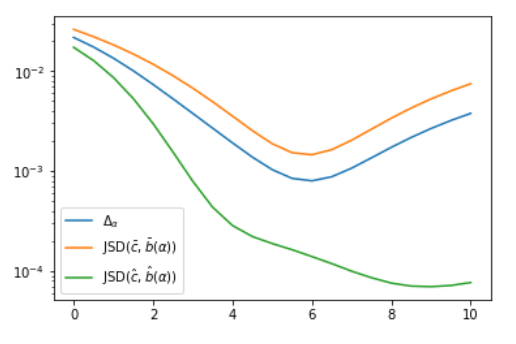}
\includegraphics[width=7.5cm]{LFR35a-best.png}
}
\vspace{1cm}
\hbox{
\includegraphics[width=5.4cm]{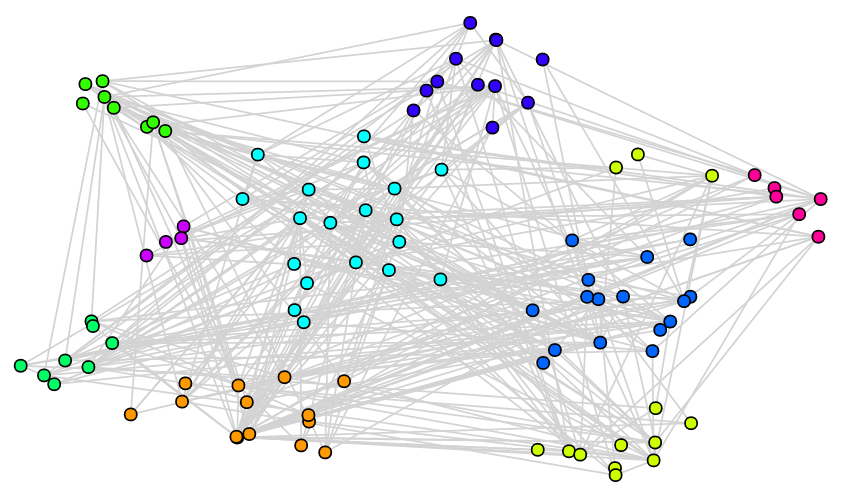}
\includegraphics[width=5.4cm]{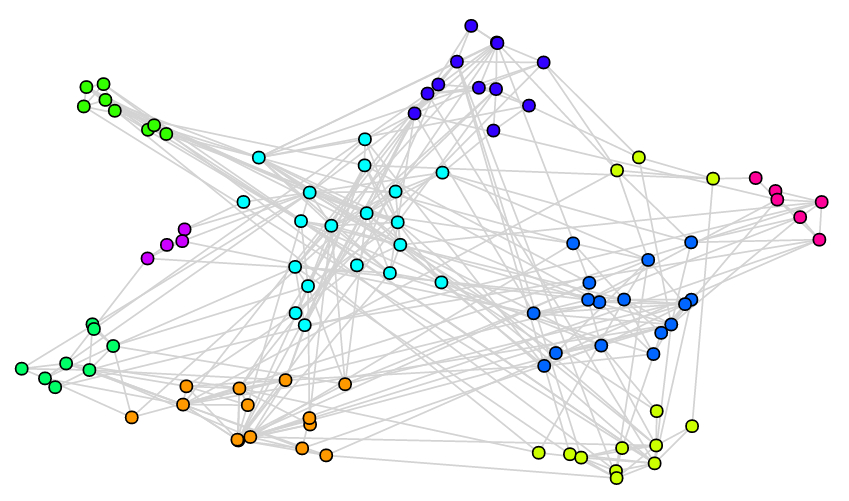}
\includegraphics[width=5.4cm]{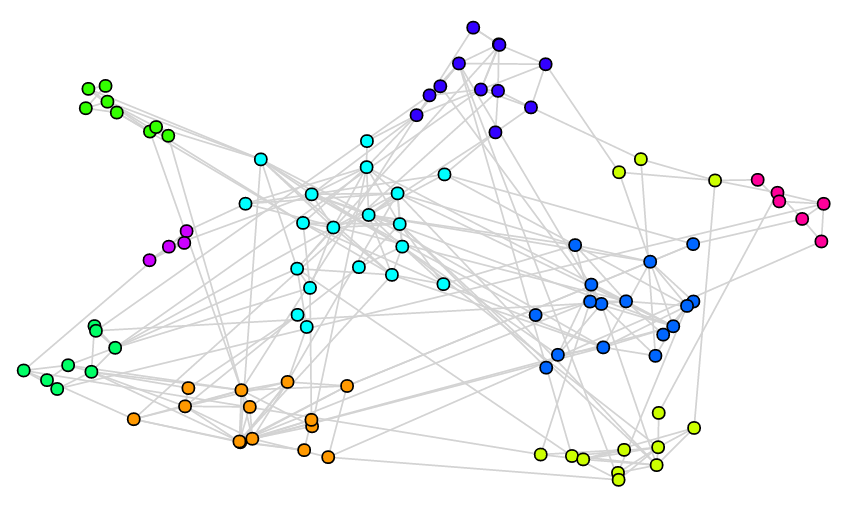}
}
\end{center}
\caption{For the LFR graph with mixing parameter $\mu=.35$, on the top row, we illustrate the divergence score as a function of $\alpha$ (left) for the best embedding found by our framework (right).
On the bottom row, we generate edges via the
Geometric Chung-Lu Model with distance parameters $\alpha=0$ (left),
the optimal value $\hat{\alpha}=6$ (center) and $\alpha=10$ (right).
\label{fig:lfr35a-alpha}}
\end{figure}

Results for the LFR graph with mixing parameter $\mu=0.35$ are shown in Figure \ref{fig:lfr35a-alpha}.
For the best embedding, vertices within communities are not as close as in the previous LFR graph, as expected. Conclusions regarding the values of $\alpha$ are the same as for the previous examples.

\clearpage


\subsection{Noisy LFR Graph}

\begin{figure}[ht]
\begin{center}
\hbox{
\includegraphics[width=7.5cm]{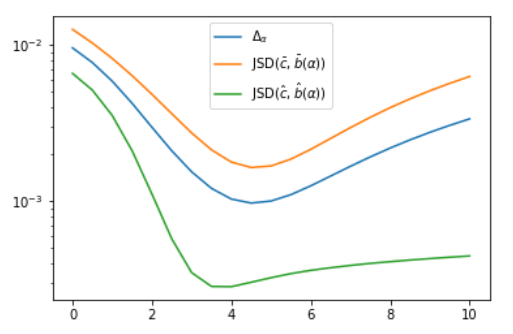}
\includegraphics[width=7.5cm]{LFR55a-best.png}
}
\vspace{1cm}
\hbox{
\includegraphics[width=5.4cm]{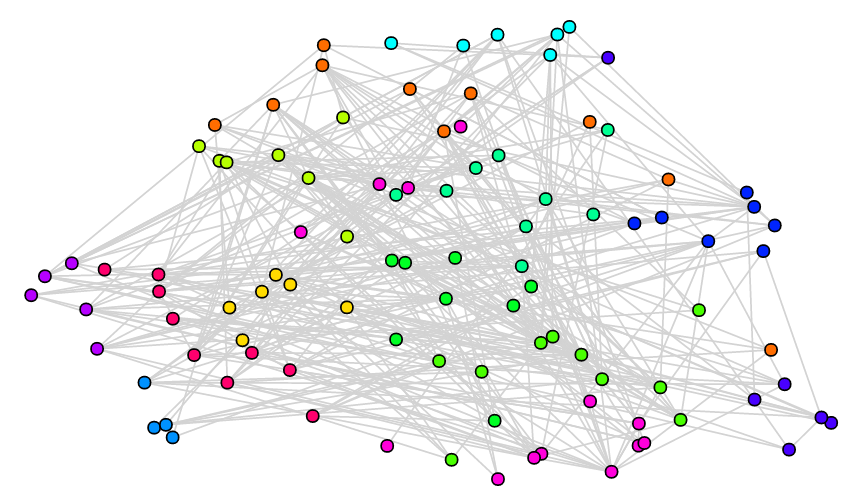}
\includegraphics[width=5.4cm]{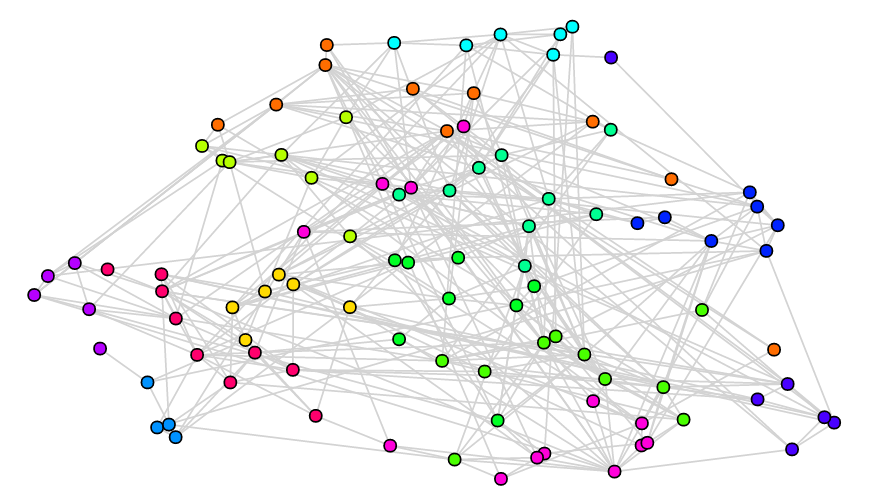}
\includegraphics[width=5.4cm]{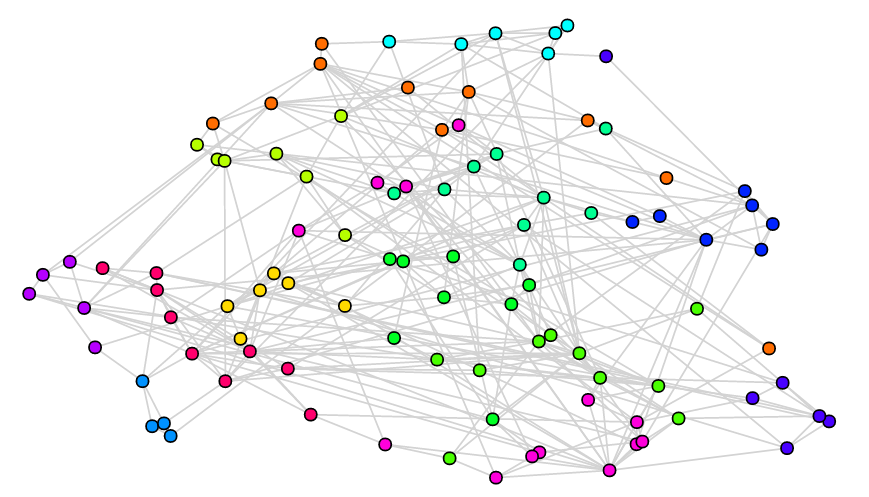}
}
\end{center}
\caption{For the LFR graph with mixing parameter $\mu=0.55$, on the top row, we illustrate the divergence score as a function of $\alpha$ (left) for the best embedding found by our framework (right).
On the bottom row, we generate edges via the
Geometric Chung-Lu Model with distance parameters $\alpha=0$ (left),
the optimal value $\hat{\alpha}=4.5$ (center), and $\alpha=7$ (right).
\label{fig:lfr55a-alpha}}
\end{figure}

In this graph, LFR graph with mixing parameter $\mu=0.55$, for each community, there are more expected external edges than internal edges. Nevertheless, in the best embedding shown in Figure~\ref{fig:lfr55a-alpha}, we still see some spatial grouping of the communities. 

\begin{thebibliography}{10}

\bibitem{Newman_book} Newman~M. Networks: An Introduction. Oxford University Press; 2010.

\bibitem{Bianconi2015} Bianconi~G. Interdisciplinary and physics challenges of network theory. EPL. 2015; 111(5):56001.

\bibitem{Hoff2002} Hoff PD, Raftery AE, Handcock MS. Latent space approaches to social network analysis,
J.Amer.Stat.Assoc. 2002; 97(460) 1090-1098.

\bibitem{Krioukov2016} Krioukov D. Clustering means geometry in networks. Phys Rev Lett. 2016; 208302(May):1-5.

\bibitem{Zuev2015} Zuev K, Boguna M, Bianconi G, Krioukov D. Emergence of Soft Communities from Geometric Preferential Attachment. Scientific Reports. 2015; 5,9421.

\bibitem{Gastner2006} Gastner MT, Newman MEJ. The spatial structure of networks. European Physical Journal B. 2006; 49(2):247-252.

\bibitem{Expert2011} Expert P, Evans TS, Blondel VD, Lambiotte R. Uncovering space-independent communities in spatial networks. Proceedings of the National Academy of Sciences. (2011); 108(19):7663-7668.

\bibitem{Janssen2010} J.~Janssen. Spatial Models for Virtual Networks. CiE 2010, LNCS 6158, pp.~201-210, 2010.

\bibitem{Higham2008} Higham DJ, Rasajski M, Przulj
N. Fitting a geometric graph to a protein-protein interaction network. Bioinformatics. 2008; 24(8):1093-1099.

\bibitem{Poulin2019} Poulin V., Th\'eberge F. (2019) Ensemble Clustering for Graphs. In: Aiello L., Cherifi C., Cherifi H., Lambiotte R., Li\'o P., Rocha L. (eds) Complex Networks and Their Applications VII. COMPLEX NETWORKS 2018. Studies in Computational Intelligence, vol 812. Springer, Cham.

\bibitem{Lin1991} Lin~J. Divergence measures based on the shannon entropy. In: IEEE Transactions
on Information theory, 37(1), pp.145-151 (1991)

\bibitem{CL} Chung~FRK, Lu~L. Complex Graphs and Networks. American Mathematical Society; 2006.

\bibitem{Seshadhri2012}  Seshadhri~C,  Kolda~TG, Pinar~A.  Community structure and scale-free collections of Erd\"os-R\'enyi graphs. Physical Review E. 2012; 85: 056109.

\bibitem{Kolda2014} Kolda~TG, Pinar~A, Plantenga~T, Seshadhri~C. A scalable generative graph model with community  structure. SIAM Journal on  Scientific  Computing. 2014; 36: C424--C452.

\bibitem{Winlaw2015} Winlaw~M, DeSterck~H, Sanders~G. An In-Depth Analysis of the Chung-Lu Model. Lawrence Livermore Technical Report LLNL-TR-678729. 2015; doi: 10.2172/1239211.

\bibitem{Lu2018} Z.\ Lu, J.\ Wahlström, A.\ Nehorai, Community Detection in Complex Networks via Clique Conductance. Nature SCiENtifiC REPORTS (2018) 8:5982.

\bibitem{LFR} A.~Lancichinetti, S.~Fortunato, and F.~Radicchi. Benchmark graphs for testing community detection algorithms. Phys.\ Rev.\ E, 78(4), 2008.

\bibitem{Zachary77} Zachary, W. W. An information flow model for conflict and fission in small groups. Journal of Anthropological Research 33, 452-473 (1977).

\bibitem{Football} M.\ Girvan, M.E.\ Newman. Community structure in social and biological networks. Proceedings of the National Academy of Sciences 99, 7821-7826 (2002).

\bibitem{snapnets} J.\ Leskovec, A.\ Krevl. SNAP Datasets: Stanford Large Network Dataset Collection, \url{http://snap.stanford.edu/data}.

\bibitem{Barabasi2016} A.L.\ Barabasi. Network Science. Cambridge U Press, 2016.

\bibitem{node2vec} A.\ Grover, J.\ Leskovec. node2vec: Scalable Feature Learning for Networks. KDD 2016: 855--864.

\bibitem{verse} A.\ Tsitsulin, D.\ Mottin, P.\ Karras, and E.\ Müller. 2018. VERSE: Versatile Graph Embeddings from Similarity Measures. In Proceedings of the 2018 World Wide Web Conference (WWW'18). International World Wide Web Conferences Steering Committee, Republic and Canton of Geneva, Switzerland, 539-548.

\bibitem{LINE} J.  Tang,  M.  Qu,  M.  Wang,  M.  Zhang,  J.  Yan,  Q.  Mei,  Line:  Large-
scale  information  network  embedding,  in:   Proceedings  24th  International
Conference on World Wide Web, 2015, pp. 1067–1077.

\bibitem{ECG} Poulin V., Th\'eberge F. Ensemble Clustering for Graphs:Comparisond and Applications., pre-print, arXiv:1903:08012, 2019.

\bibitem{louvain} Blondel, V. D., Guillaume, J.-L., Lambiotte, R. and Lefebvre, E. Fast unfolding of communities in large networks. Journal of Statistical Mechanics: Theory and Experiment 2008, P10008 (2008).

\bibitem{infomap} Rosvall M, and Bergstrom C.T., Maps of random walks on complex networks reveal community structure, PNAS 105 (1118), 2008.

\bibitem{UMAP} L.\  McInnes, J.\ Healy, J.\  Melville. UMAP: Uniform Manifold Approximation and Projection for Dimension Reduction. pre-print arXiv:1802.03426, 2018.

\bibitem{Modularity} B.\ Ka\'minski, V.\ Poulin, P.\ Pra\l{}at, P.\ Szufel, and F.\ Th\'eberge, Clustering via Hypergraph Modularity, preprint.

\bibitem{Hypergraphs} A.\ Antelmi, G.\ Cordasco, B.\ Kami\'nski, P.\ Pra\l{}at, V.\ Scarano, C.\ Spagnuolo, and P.\ Szufel, SimpleHypergraphs.jl --- Novel Software Framework for Modelling and Analysis of Hypergraphs, Proceedings of the 16th Workshop on Algorithms and Models for the Web Graph (WAW 2019), Lecture Notes in Computer Science 11631, Springer, 2019.

\end{thebibliography}
\end{document}